\documentclass[wcp]{jmlr}

\usepackage{longtable}

\usepackage{booktabs}
\usepackage{multirow}

\makeatletter
\let\Ginclude@graphics\@org@Ginclude@graphics 
\makeatother

\jmlryear{2022}
\jmlrworkshop{ACML 2022}
\jmlrvolume{189}

\title[AS-IntroVAE]{AS-IntroVAE: Adversarial Similarity Distance Makes Robust IntroVAE}

  \author{\Name{Changjie Lu} \Email{lucha@kean.edu}\\
  \addr Wenzhou-Kean University, Wenzhou, China
  \AND
  \Name{Shen Zheng} \Email{shenzhen@andrew.cmu.edu}\\
  \addr Carnegie Mellon University, Pittsburgh, USA\\
  Wenzhou-Kean University, Wenzhou, China
  \AND
  \Name{Zirui Wang} \Email{lx1936@126.com}\\
  \addr Zhejiang University, Hangzhou, China
  \AND
  \Name{Omar Dib} \Email{odib@kean.edu} \\
  \addr Wenzhou-Kean University, Wenzhou, China
  \AND
    \Name{Gaurav Gupta} \Email{ggupta@kean.edu}\\
  \addr Wenzhou-Kean University, Wenzhou, China
 }

\begin{document}

\maketitle

\begin{abstract}


Recently, introspective models like IntroVAE and S-IntroVAE have excelled in image generation and reconstruction tasks. The principal characteristic of introspective models is the adversarial learning of VAE, where the encoder attempts to distinguish between the real and the fake (i.e., synthesized) images. However, due to the unavailability of an effective metric to evaluate the difference between the real and the fake images, the posterior collapse and the vanishing gradient problem still exist, reducing the fidelity of the synthesized images. In this paper, we propose a new variation of IntroVAE called Adversarial Similarity Distance Introspective Variational Autoencoder (AS-IntroVAE). We theoretically analyze the vanishing gradient problem and construct a new Adversarial Similarity Distance (AS-Distance) using the 2-Wasserstein distance and the kernel trick. With weight annealing on AS-Distance and KL-Divergence, the AS-IntroVAE are able to generate stable and high-quality images. The posterior collapse problem is addressed by making per-batch attempts to transform the image so that it better fits the prior distribution in the latent space. Compared with the per-image approach, this strategy fosters more diverse distributions in the latent space, allowing our model to produce images of great diversity. Comprehensive experiments on benchmark datasets demonstrate the effectiveness of AS-IntroVAE on image generation and reconstruction tasks.





\end{abstract}



\section{Introduction}

    In the last decade, two types of deep generative models---Variational Autoencoders (VAEs) (\cite{kingma2013auto}), and Generative Adversarial Networks (GANs) (\cite{goodfellow2014generative}) --- has gained tremendous popularity in computer vision (CV) applications. Their acceptance is attributed to their success in various CV tasks, such as image generation (\cite{karras2019style, vahdat2020nvae}), image reconstruction (\cite{gu2020image, hou2017deep}), and image-to-image translation (\cite{zhu2017unpaired, liu2017unsupervised}). 
    
    VAEs can produce images with diverse appearances and is easy-to-train. However, the synthesized images of VAEs are often blurry and lack fine details (\cite{larsen2016autoencoding}). On the other hand, GANs produce sharper images with more details but often suffer from mode collapse (i.e., lack diversity) and vanishing gradient (\cite{goodfellow2016nips}). Many researchers have sought to develop an efficient hybrid model that combines the advantages of VAEs and GANs. Unfortunately, due to the requirement of an extra discriminator, existing hybrid VAE and GAN models (\cite{makhzani2015adversarial, larsen2016autoencoding, dumoulin2016adversarially, tolstikhin2017wasserstein}) has high computational complexity and heavy memory usage. Even with these delicate architecture designs, those methods still underperform leading GANs (\cite{karras2017progressive, brock2018large}) in terms of the quality of the generated images.
    
    
    Unlike classical hybrid GAN-VAE models, introspective methods (\cite{huang2018introvae, daniel2021soft}) eliminate the need for extra discriminators. Instead, they utilize the decoder as the `actual' discriminator to distinguish between the fake and the real images and have achieved state-of-the-art results on image generation tasks. Despite their progress, those introspective learning-based methods suffer from the \textbf{posterior collapse} problem with insufficiently tuned hyperparameters and the \textbf{vanishing gradient} problem, especially during the early-stage model training.
    

    
    In this paper, we propose Adversarial Similarity Distance Introspective Variational Autoencoder (AS-IntroVAE), an introspective VAE that can competently address the posterior collapse and the vanishing gradient problem. Firstly, we present a theoretical analysis and demonstrate that the vanishing gradient problem of introspective models can be addressed by a similarity distance based upon the 2-Wasserstein distance and the kernel trick. We termed this distance as Adversarial Similarity Distance (AS-Distance). The weight annealing strategy applied to the AS-Distance and the KL-Divergence yields highly stable synthesized images with excellent quality. We address the posterior collapse problem by per-batch aligning the image with the prior distribution in the latent space. This strategy allows the proposed AS-IntroVAE to contain diverse distributions in the latent space, thereby promoting the diversity of the synthesized image.

    
    
    Our contribution are highlighted as follows (1) A new introspective variational autoencoder named Adversarial Similarity Distance Introspective Variational Autoencoder (AS-IntroVAE) (2) A new theoretical understanding of the posteriors collapse and the vanishing gradient problem in VAEs. (3) A novel similarity distance named Adversarial Similarity Distance (AS-Distance) for measuring the differences between the real and the synthesized images. (4) Promising results on image generation and image reconstruction tasks with significantly faster convergence speed.

    
    
        

\section{Related Work}

\subsection{Generative Adversarial Network (GAN)}

Generative Adversarial Network (GAN) (\cite{goodfellow2014generative}) consists of a generator $G(z)$ and a discriminator $D(x)$. The generator tries to confuse the discriminator by generating a synthetic image from the input noise $z$, whereas the discriminator tries to distinguish that synthetic image from the real image $x$. 



There are two crucial drawbacks with vanilla GANs: mode collapse (insufficient diversity) and vanishing gradient (insufficient stability) (\cite{goodfellow2016nips}). To remedy these issues, WGAN (\cite{arjovsky2017wasserstein}) replace the commonly used Jensen Shannon divergence with Wasserstein distance to alleviate vanishing gradient and mode collapse. However, the hard weight clipping used by WGAN to satisfy the Lipschitz constraint significantly reduces the network capacity (\cite{gulrajani2017improved}). Some better substitutes for weight clipping include gradient penalty of WGAN-GP (\cite{gulrajani2017improved}) and spectral normalization (\cite{miyato2018spectral}) of SN-GAN.


Compared with these GAN approaches, our method also enjoys the advantages of VAE-based methods: more diversity and stability for the synthesized images.






\subsection{Variational Autoencoder (VAE)}

Variational Autoencoder (VAE) (\cite{kingma2013auto}) consists of an encoder and an decoder. The encoder $q_{\phi}(z \mid x)$ compress the image $x$ into a latent variable $z$, whereas the decoder $p_{\theta}(x \mid z)$ try to reconstruct the image from that latent variable. Besides, variational inference is applied to approximate the posterior. 


A common issue during VAE training is posterior collapse (\cite{bowman2015generating}). Posterior collapse occurs when the latent variables become uninformative (e.g., weak, noisy) such that the model choose to rely solely on the autoregressive property of the decoder and ignore the latent variables (\cite{subramanian2018towards}). Recent approaches mainly address the posterior collapse problems using KL coefficient annealing (\cite{bowman2015generating, fu2019cyclical}), auxiliary cost function (\cite{alias2017z}), pooling operations (\cite{long2019preventing}), variational approximation restraints (\cite{razavi2019preventing}), or different Evidence Lower Bound (ELBO) learning objectives (\cite{havrylov2020preventing}).


Compared with former VAE approaches, our method also shares the strength of GAN-based models: sharp edges and sufficient fine details.

\subsection{Integration of VAE and GAN}

A major limitation of VAEs is that they tend to generate blurry, photo-unrealistic images (\cite{dosovitskiy2016generating}). One popular approach to alleviate this issue is to integrate GAN's adversarial training directly into VAE to obtain sharp edges and fine details. Specifically, these hybrid models often consists of an encoder-decoder and an extra discriminator. For example, VAE-GAN (\cite{larsen2016autoencoding}) and A-VAE (\cite{mescheder2017adversarial}) both utilize a VAE-like encoder-decoder structure and an extra discriminator to constraint the latent space with adversarial learning. ALI (\cite{dumoulin2016adversarially}) and BiGANs (\cite{donahue2016adversarial}) adopt both mapping and inverse mapping with an extra discriminator to determine which mapping result is better.



To save the growing computational cost from the extra discriminators, the recent state-of-the-art image synthesis method IntroVAE (\cite{huang2018introvae}) proposes to train VAEs in an introspective way such that the model can distinguish between the fake and real images using only the encoder and the decoder. The problem with IntroVAE is that it utilizes a hard margin to regulate the hinge terms (i.e., the KL divergence between the posterior and the prior), which leads to unstable training and difficult hyperparameter selection. To alleviate this issue, S-IntroVAE (\cite{daniel2021soft}) expresses VAE's ELBO in the form of a smooth exponential loss, thereby replacing the hard margin with a soft threshold function. However, the posterior collapse and the vanishing gradient problem still exist in these introspective methods.


Unlike these methods, our approach has stable training throughout the entire training stage and can generate samples of sufficient diversity without careful hyperparameter tuning.

\section{Background}       
We place our generative model under the variational inference setting (\cite{kingma2013auto}), where we aim to utilize variational inference methods to approximate the intractable maximum-likelihood objective. With this in mind, in this section, we will first revisit vanilla VAE, focusing on its ELBO technique, and then analyze introspective learning-based methods, including IntroVAE and S-IntroVAE.


\subsection{Evidence Lower Bound (ELBO)}
The learning object of VAE is to maximize the evidence lower bound (ELBO) as below:

\begin{equation}
\log _{\theta}(x) \geq E_{q_{\phi}(z \mid x)} \log p_{\theta}(x \mid z)-D_{K L}\left(q_{\phi}(z \mid x) \| p_{\theta}(z)\right)
\end{equation}
where $x$ is the input data, $z$ is the latent variable, $q_{\phi}(z \mid x)$ represents the encoder with parameter $\phi$, and $p_{\theta}(x \mid z)$ represents the decoder with parameter $\theta$. The Kullback-Leibler (KL) divergence term can be expressed as:

\begin{equation}
D_{\mathrm{KL}}(q(z \mid x) \| p(z))=\mathbb{E}_{q(z)}\left[\log \frac{q(z \mid x)}{p(z)}\right]
\end{equation}

Reparameterization trick (\cite{kingma2013auto}) is applied to make VAE trainable (i.e., differentiable through backpropagation). Specifically, the reparameterization trick transforms the latent representation $z$ into two latent vectors $\mu$ and $\sigma$ and a random vector $\varepsilon$, thereby excluding randomness from the backpropagation process. The reparameterization trick can be formulated as $\mathbf{z}=\boldsymbol{\mu}+\boldsymbol{\sigma} \odot \varepsilon$.


\subsection{IntroVAE}
Unlike the vanilla VAE, which optimizes upon a single lower bound, IntroVAE (\cite{huang2018introvae}) incorporates an adversarial learning strategy commonly used by GANs during training. Specifically, the encoder aims to maximize the KL divergence between the fake image and the latent variable and minimize the KL divergence between the actual image and the latent variable. Meanwhile, the decoder aims to confuse the encoder by minimizing the KL divergence between the fake photo and the latent variable. The learning objective (i.e., loss function) of IntroVAE for Encoder and Decoder is:
\begin{equation}
    \begin{aligned}
\mathcal{L}_{E} &=E L B O(x)+\sum_{s=r, g}\left[m-K L\left(q_{\phi}\left(z | x_{s}\right) \| p(z)\right]^{+}\right.\\
\mathcal{L}_{D} &=\sum_{s=r, g}\left[K L\left(q_{\phi}\left(z | x_{s}\right) \| p(z)\right)\right] .
    \end{aligned}
\end{equation}
where $x_{r}$ is the reconstructed image, $x_{g}$ is the generated image, and $m$ is the hard threshold for constraining the KL divergence.

\subsection{S-IntroVAE}
The major limitation of IntroVAE is that it utilizes a hard threshold $m$ to constrain the KL divergence term. S-IntroVAE (\cite{daniel2021soft}) suggests this design will significantly reduce model capacity and induce vanishing gradient. Instead, S-IntroVAE proposes to utilize the complete ELBO (instead of just KL) with a soft exponential function (instead of a hard threshold). The learning objective (loss function) of S-IntroVAE is:
\begin{equation}
    \begin{aligned}
&\mathcal{L}_{E}=E L B O(x)-\frac{1}{\alpha} \sum_{s=r, g} \exp \left(\alpha E L B O\left(x_{s}\right)\right) \\
&\mathcal{L}_{D}=E L B O(x)+\gamma \sum_{s=r, g} E L B O\left(x_{s}\right)
\end{aligned}
\end{equation}
where $\alpha,\gamma$ are both hyperparameters. 


\section{Proposed Method}
\begin{figure}[t]
        \centering
        \includegraphics[width=15cm]{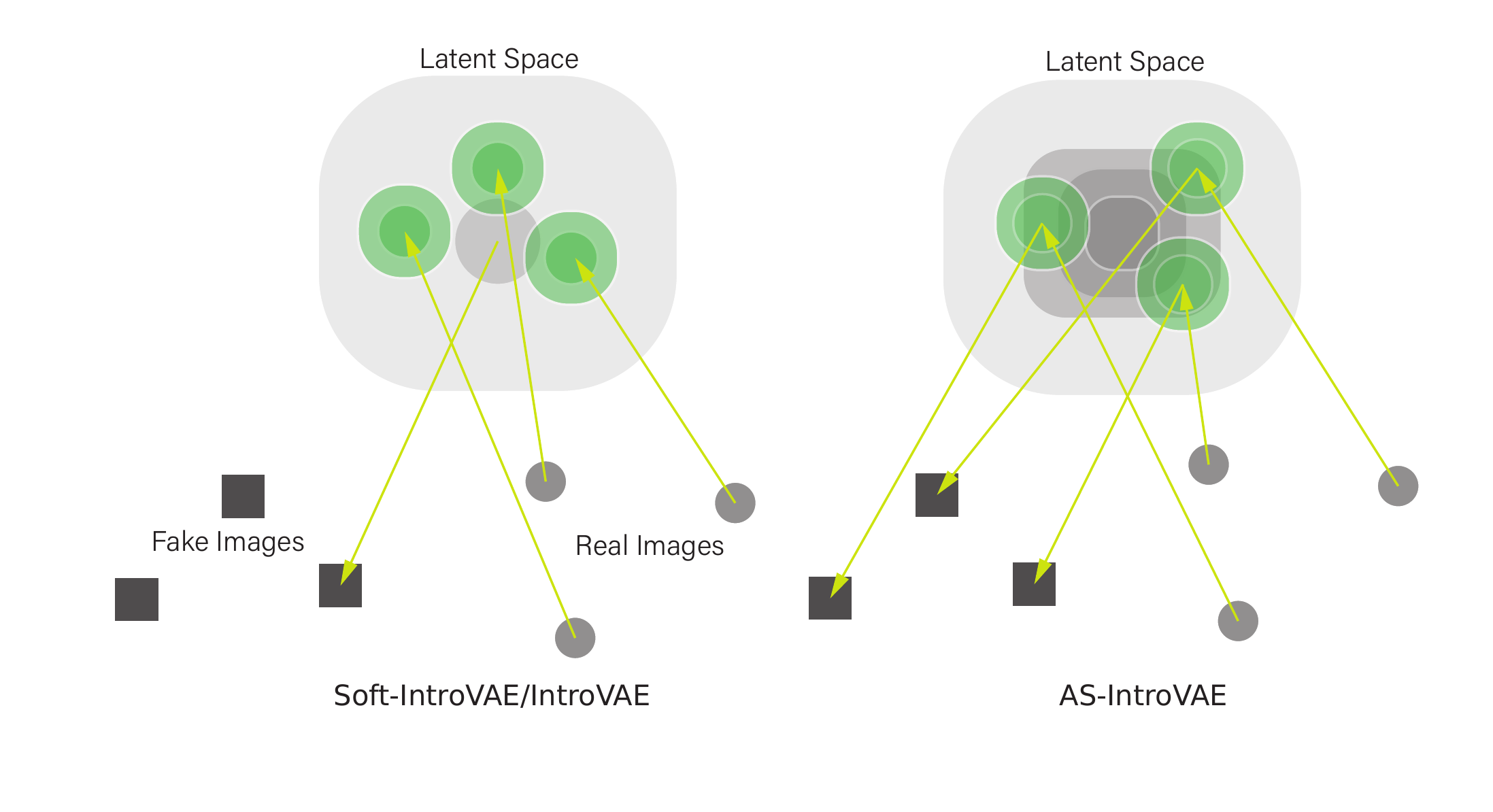}
        \caption{Illustration of how AS-IntroVAE addresses the posterior collapse problem. Both IntroVAE/S-IntroVAE and the proposed AS-IntroVAE project the real images into the latent space. However, IntroVAE/S-IntroVAE force every image to match the prior distribution of the latent space. This enforcement undermines the valuable signal from the real image such that the latent space becomes uninformative for the decoder. In contrast, AS-IntroVAE align the image with the prior distribution in a per-batch manner. Since a batch contains far more variations than a single image, in AS-IntroVAE, the signal becomes strong enough such that the decoder has to leverage the latent space to generate diverse samples.}
        \label{fig:post_collapse}
    \end{figure}

\begin{figure}[t]
        \centering
        \includegraphics[width=16cm]{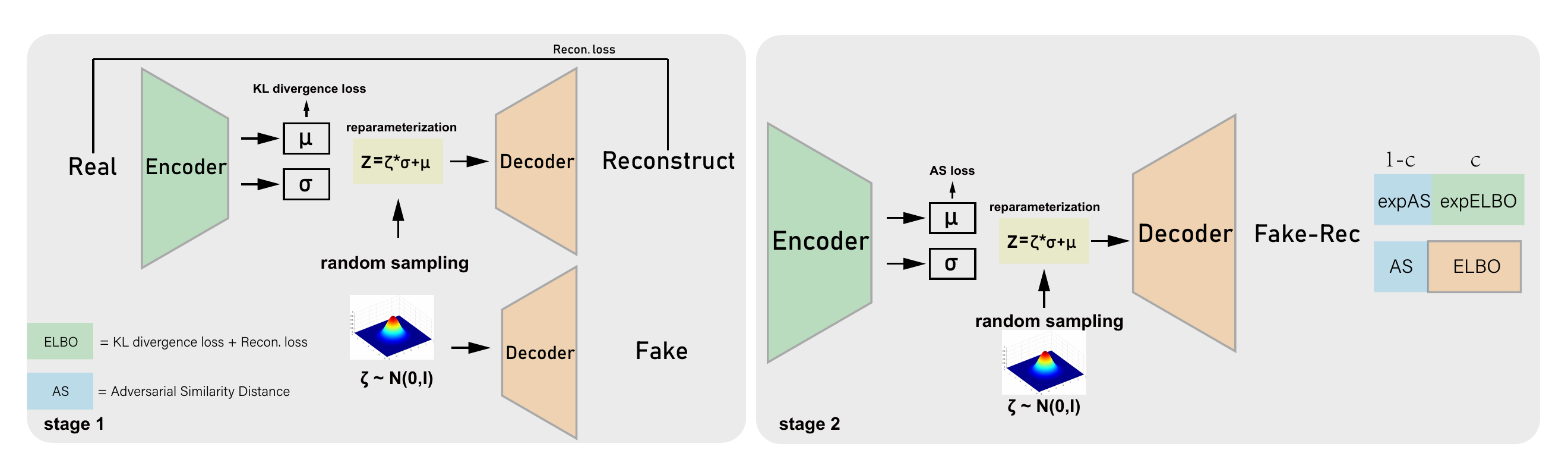}
        \caption{AS-IntroVAE workflow, using image reconstruction task as the example. AS-IntroVAE contains an encoder-decoder architecture. In the first phase, the encoder-decoder receives the real image and produce the reconstructed image. Meanwhile, the decoder will generate fake image from Gaussian noise alone. In the second phase, the \textbf{same} encoder-decoder conduct adversarial learning in the latent space for the reconstructed image and the fake image. During adversarial learning, the encoder tries to maximize the adversarial similarity distance between fake image and reconstruction image, whereas the decoder wants to minimize it. After each iteration, the model will calculate the annealing rate $c$..}
        \label{fig:workflow}
    \end{figure}

In this section, we will illustrate the proposed AS-IntroVAE, including its strategy for posterior collapse (Fig. \ref{fig:post_collapse}), its model workflow (Fig. \ref{fig:workflow}), and the theoretical analysis.

\subsection{Theoretical Analysis}
An effective distance metrics are crucial for generative model like VAEs and GANs. To address the vanishing gradient problems of S-IntroVAE and IntroVAE, we propose a novel similarity distance called Adversarial Similarity Distance (AS-Distance). Inspired by 1-Wasserstein distance (\cite{arjovsky2017wasserstein}) and the distance metrics in unsupervised domain adaptation (\cite{wu2021unsupervised}), which could provide stable gradients, the AS-Distance is defined as:
    
\begin{equation}
D\left(p_{r}, p_{g}\right) =\mathbb{E}_{x\sim p(z)}[ \left(\mathbb{E}_{x\sim p_{r}}\left[q\left(z|x\right)\right]-\mathbb{E}_{x\sim p_{g}}\left[q\left(z|x\right)\right]\right)]^{2}
\label{as}
\end{equation}
where $p_{r}$ is distribution of real data, $p_{g}$ is distribution of generated data.
The encoder and the decoder plays an adversarial game on this distance:
\begin{equation}
    arg\min_{Dec}\max_{Enc} D\left(p_{r}, p_{g}\right)
\end{equation}
We use 2-Wasserstein so that we could apply a kernel trick on Equ.\ref{as}.
\begin{equation}
D(p_{r},p_{g}) = \mathbb{E}_{x\sim p_{r,g}}\left[k\left(x_{r}^{i}, x_{r}^{j}\right)+k\left(x_{g}^{i}, x_{g}^{j}\right)-2 k\left(x_{r}^{i}, x_{g}^{j}\right)\right]
\end{equation}
where $k\left(x_{r}^{i}, x_{g}^{j}\right) = \mathbb{E}_{z\sim p(z)}[q(z|x_{r}^{i})\cdot q(z|x_{g}^{i})]$. \\
Since the latent space is a normal distribution. This kernel k can be deduced as
\begin{equation}
    k\left(x_{r}^{i}, x_{g}^{j}\right)=\frac{-\frac{1}{2} \frac{\left(u_{r}^{i}-u_{g}^{j}\right)^{2}}{\lambda_{r}^{i}+\lambda_{g}^{j}}}{(2 \pi)^{\frac{n}{2}} \cdot\left(\lambda_{r}^{i}+\lambda_{g}^{j}\right)^{\frac{1}{2}}}
\end{equation}
where $u,\lambda$ represent the variational inference on the mean and variance of $x$, $i,j$ represent the $ith, jth$ pixel in images. \\

In the maximum mean discrepancy (MMD) method, the distance calculation is conducted \textit{after} the reparameterization. As shown by (\cite{wu2020cf}), this will leads to high variance (i.e., error) for the estimated distance. Instead, we seek to calculate the distance \textit{before} the reparameterization, which can reduce the variance and improve the distance estimation accuracy.

During the experiment, we found that KL term from S-IntroVAE would generate sharp but distort images, whereas our AS term (without KL term) would generate diverse but blur images. If we fix the weight for KL and AS term (e.g. both at 0.5), there will exist two optimal solutions, which induces training stability. Inspired by (\cite{fu2019cyclical}), we decide to gradually increase the weight for KL (from 0 to 1), and decrease the weight for AS (from 1 to 0) during training. In this way, for KL and AS, we can enjoy their advantages and eschew their disadvantages.




Based on the former discussions, we derive the loss function for AS-IntroVAE as:
    \begin{equation}
        \begin{aligned}
\mathcal{L}_{E_{\phi}}&=E L B O(x)-\frac{1}{\alpha} \sum_{s=r,g} \exp (\alpha( \mathbb{E}_{q(z \mid x_{s})}[\log p(x \mid z)] \\
&+c K L(q_{\phi}(z|x_{s}) \| p(z))+(1-c) D(p_{r}, p_{g}))) \\
\mathcal{L}_{D_{\theta}}&=E L B O(x)+\gamma \sum_{s=r,g}( \mathbb{E}_{q(z \mid x_{s})}[\log p(x \mid z)]\\
& +c K L(q_{\phi}(z|x_{s}) \| p(z))+(1-c) D(p_{r},p_{g}) )
\end{aligned}
\label{loss}
    \end{equation}
where $c = min(i*5/T,1)$, $i$ is the current iteration and $T$ is total iteration.
    
\begin{theorem}
    Introspective Variational Autoencoders (IntroVAEs) have vanishing gradient problems.
\end{theorem}
\begin{proof}
As illustrated in IntroVAEs (IntroVAE and S-IntroVAE), the Nash equilibrium can be attained when
$K L\left(q_{\phi}\left(z|x_{r}\right) \| q_{\phi}\left(z|x_{g}\right)\right) = 0$, where $x_{r}$ could also represents the real images since the reconstructed images are sampled from real data points. Moreover, with the object $D_{K L}\left(q_{\phi}(z \mid x) \| p(z)\right) = 0$, we have:
\begin{equation}
    q_{\phi}\left(z|x_{r}\right)=q_{\phi}\left(z|x_{g}\right)=p(z)
\end{equation}
Replace the term $p(z)$ with $\frac{q_{\phi}\left(z|x_{r}\right)+q_{\phi}\left(z|x_{g}\right)}{2}$, the adversarial term for the decoder then becomes:
\begin{equation}
\begin{aligned}
    &K L\left(q_{\phi}\left(z|x_{r}\right) \| \frac{q_{\phi}\left(z|x_{r}\right)+q_{\phi}\left(z|x_{g}\right)}{2}\right)+K L\left(q_{\phi}\left(z|x_{g}\right) \| \frac{q_{\phi}\left(z|x_{r}\right)+q_{\phi}\left(z|x_{g}\right)}{2}\right) \\
    & =2 J S D\left(q_{\phi}\left(z|x_{r}\right) \| q_{\phi}\left(z|x_{g}\right)\right)
\end{aligned}
\end{equation}
\end{proof}
Therefore, the gradient of loss for Decoder in IntroVAE becomes:
\begin{equation}
     \nabla \mathcal{L}_{D} = \nabla 2 J S D\left(q_{\phi}\left(z|x_{r}\right) \| q_{\phi}\left(z|x_{g}\right)\right)
\end{equation}
As shown by (\cite{arjovsky2017towards}), if $P_{x_{r}}$ and $P_{x_{g}}$ are two distributions in two different manifolds that don't align perfectly and don't have full dimension (i.e., the dimension of the latent variable is sparse in the image dimension). With the assumption that $P_{x_{r}}$ and $P_{x_{g}}$ are continuous in their manifolds, if there exists a set A with measure 0 in one manifold, then $P(A) = 0$. Consequently, there will be an optimal discriminator with 100$\%$ accuracy for classify almost any $x$ in these two manifolds, resulting in $\nabla \mathcal{L}_{D} = 0$.

For IntroVAE, the above condition (i.e., vanishing gradient problem) occurs since the very beginning of the training process when there is no intersection between the real image distribution and the fake image distribution. The reason why S-IntroVAE alleviate vanishing gradient at the later epoch (See Fig.\ref{fig:train_stability}) is that the reconstruction loss of S-IntroVAE gradually create a support set between the real and the fake images during training. In comparison, since AS-Distance is based on 2-Wasserstein distance, the proposed AS-IntroVAE provides stable gradient, even there is no intersection between these two distributions.
    


    
    
    

\section{Experiments}

In this section, we will explain the implementation details, the comparison on 2D toy datasets, image generation and image reconstruction tasks, and the training stability. 

    \subsection{Implementation Details}
    
    We train our model using the Adam (\cite{kingma2014adam}) optimizer with the default setting ($\beta_{1}=0.9$ and $\beta_{2}=0.999$) for 150 epochs. We implement our framework in Pytorch (\cite{paszke2019pytorch}) with 3 NVIDIA RTX 3090 GPU. Same with (\cite{huang2018introvae, daniel2021soft}), we set a fixed learning rate of 2e-4. It takes around 1 day to converge our model on CelebA-128 dataset and 2 days on CelebA-256 dataset, respectively. The exponential moving average is applied to stabilize training. The encoder and decoder will be updated respectively in each iteration. For the loss function, we set $\alpha = 2$, and $\gamma = 1$. The weight for the real image's KL term and reconstruction term is fixed at 0.5 and 1.0, respectively, whereas the fake image's KL term and reconstruction term is both set at 0.5. For the annealing rate $c$, we apply a linear function shown in Equ.\ref{loss}. For other hyperparameter settings, we inherit the setting from S-IntroVAE (\cite{daniel2021soft}).
    
        

    




    \subsection{2D Toy Datasets}

    In this subsection, we evaluate the proposed method's performance on 2D Toy datasets, including Gaussian and Checkerboard (\cite{de2020block, grathwohl2018ffjord}), and compare our approach with baselines including VAE, IntroVAE, and S-IntroVAE using two commonly used metrics, including KL-divergence (KL) and Jensen–Shannon-divergence (JSD). Both KL and JSD measure how far away the model predicted outcome is from the ground truth data distribution. Therefore, a lower score indicates a better result for both KL and JSD.
        
    We design three hyperparameter combinations to assess the robustness of different methods to hyperparameter changes. The hyperparameter includes the weight for real image ELBO, the weight for fake image's KL divergence term, and the weight for fake image's reconstruction term. The value for each combination is as below. \textit{C1: (0.3, 0.1, 0.9) C2: (0.5, 0.1, 0.9) C3: (0.7, 0.2, 0.9)}.
    
    Table \ref{tab:toy_dataset} shows the quantitative comparison on 2D Toy Datasets 8 Gaussians.  For the 8 Gaussians dataset, we find that our method has the lowest (i.e., best) KL and JSD score under all hyperparameter combinations, outperforming VAE, IntroVAE, and S-IntroVAE by a large margin.
    
\begin{figure}[t]
        \centering
        \includegraphics[width=3cm]{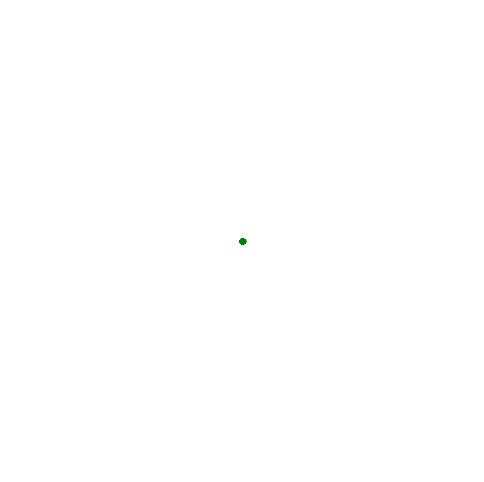}
        \includegraphics[width=3cm]{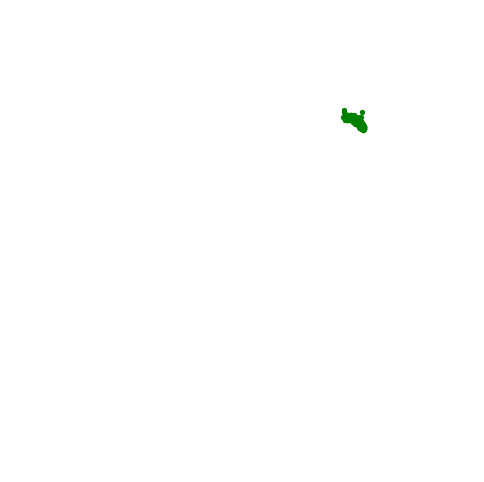}
        \includegraphics[width=3cm]{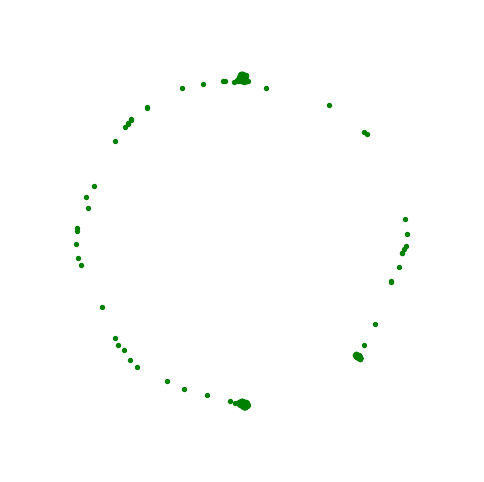}
         \includegraphics[width=3cm]{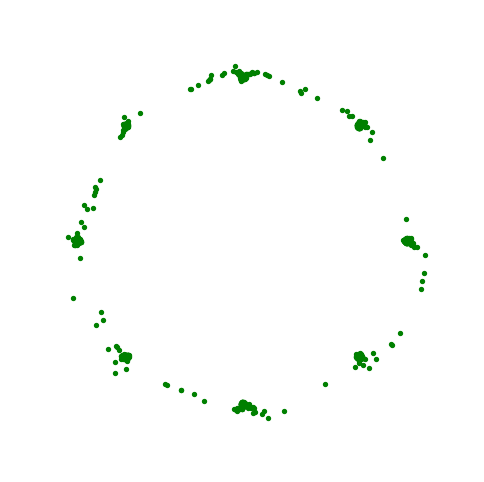}

        \includegraphics[width=3cm]{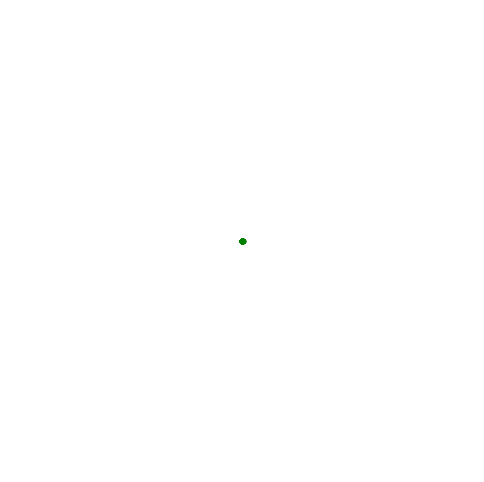}
        \includegraphics[width=3cm]{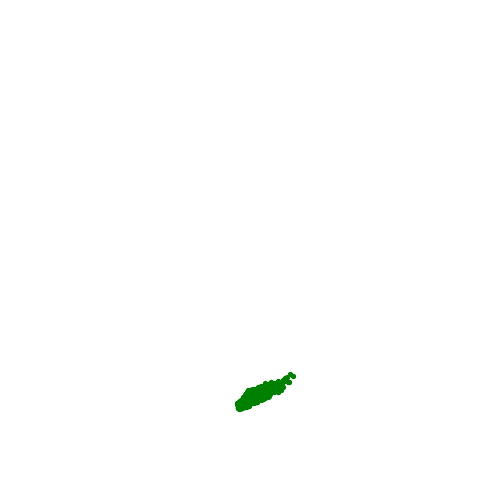}
        \includegraphics[width=3cm]{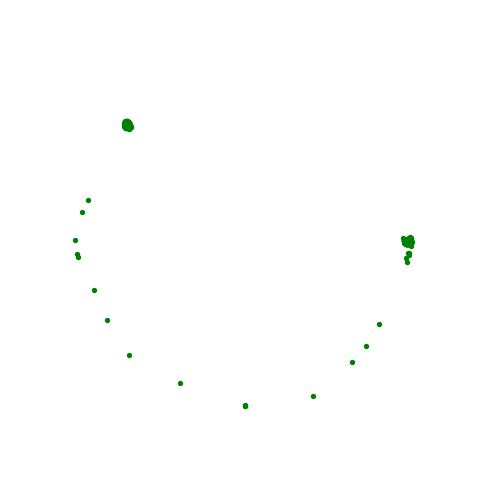}
         \includegraphics[width=3cm]{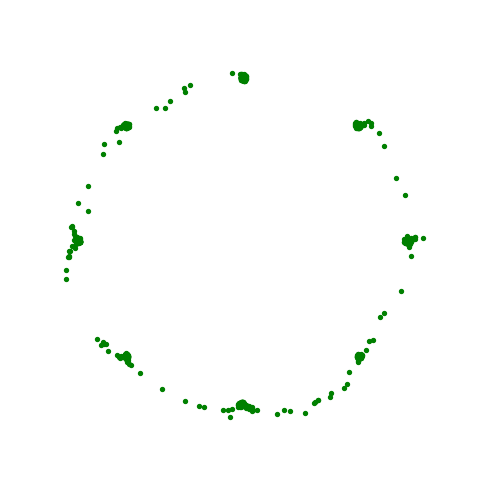}
        
        \includegraphics[width=3cm]{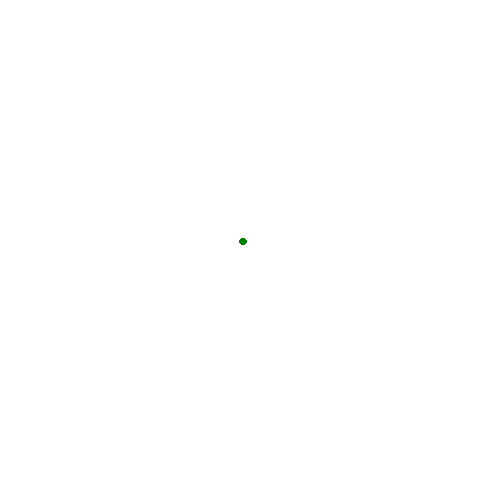}
        \includegraphics[width=3cm]{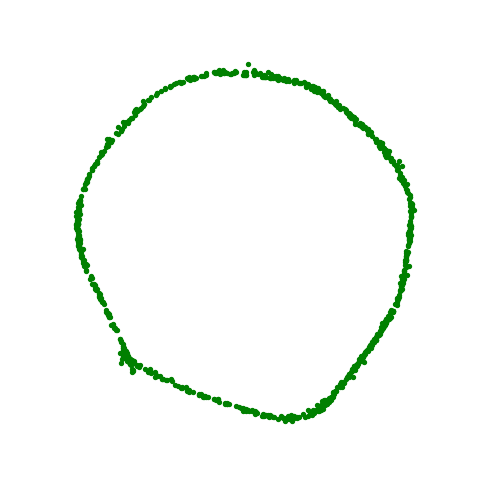}
        \includegraphics[width=3cm]{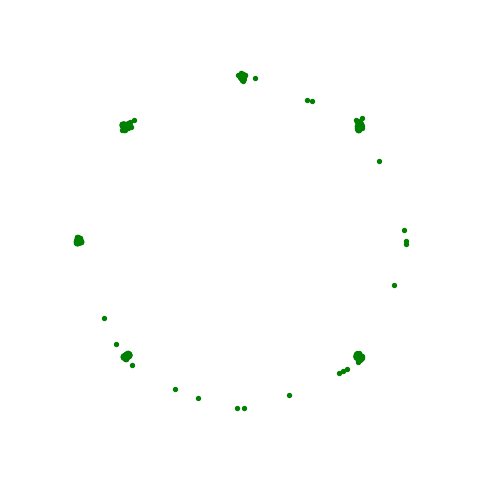}
        \includegraphics[width=3cm]{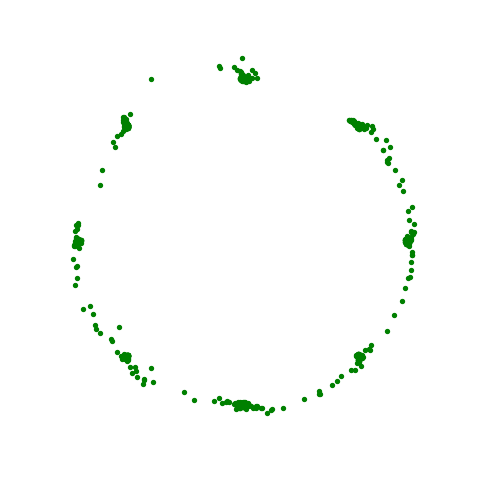}  
       
        \caption{Visual Comparison on 2D Toy Dataset 8 Gaussians. From top to bottom row: results with different hyperparameters. From left to right column: VAE, IntroVAE, S-IntroVAE, Ours. Zoom in to view the detail within each subfigure.}
        \label{fig:8Gaussians}
    \end{figure}

    Fig. \ref{fig:8Gaussians} shows the qualitative comparison of the 8 Gaussians dataset. We notice that VAE has one isolated data point for all three hyperparameter combinations, which means it has severe posterior collapse problems. IntroVAE has a small trace around a specific data point for C1 and C2, indicating nontrivial posterior collapse problems. For C3, IntroVAE produces a ring shape, meaning the generated data is evenly distributed and fails to converge to any designated data point. S-IntroVAE converges to two data points for C1 and C2 and six for C3, which still reflects the posterior collapse problem.

    In comparison, our method, under all hyperparameter combinations, successfully converges to all eight data points. Therefore, we can conclude our approach is the only one that avoids the posterior collapse problem in 8 Gaussian datasets experiments. Due to the scope of this paper, the result for the Checkerboard dataset is in the supplementary material.

    \subsection{Image Generation}
    In this subsection, we evaluate the proposed method's performance on image generation tasks, using benchmark datasets including MNIST (\cite{lecun1998gradient}), CIFAR-10 (\cite{krizhevsky2009learning}), CelebA-128, and CelebA-256 (\cite{liu2015deep}). The methods for comparison includes WGAN-GP and S-IntroVAE, and the evaluation metric is Frechet Inception Distance (FID). Specifically, we use FID to estimate the distance between the generated dataset's distribution and the source (i.e., training) dataset's distribution. Hence, a lower FID score means a better result. 
    
\begin{figure}[t]
    \centering
        \subfigure[WGAN-GP]{
            \includegraphics[width=4.7cm]{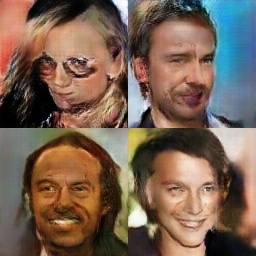}
        }
        \subfigure[S-IntroVAE]{
            \includegraphics[width=4.7cm]{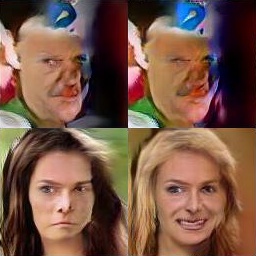}
        }
        \subfigure[Ours]{
            \includegraphics[width=4.7cm]{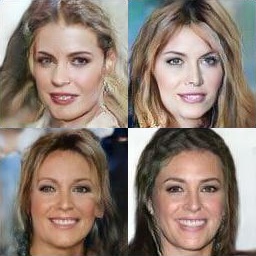}
        } 
    \caption{Image Generation Visual Comparison at CelebA-128 dataset.}
    \label{fig:gen_celeba_128}
    \end{figure}

    Table \ref{tab:fid} shows the quantitative comparison of image generation tasks. For all chosen datasets, the proposed method has the lowest (e.g., best) FID score. Fig. \ref{fig:gen_celeba_128} shows the qualitative comparison of image generation at CelebA-128 dataset. We notice both WGAN-GP and S-IntroVAE have apparent facial feature distortion, edging blur, and facial asymmetry. Although S-IntroVAE has less ghost artifact and unnatural texture than WGAN-GP, it has a significant posterior collapse problem: S-IntroVAE's two generated images in the first row are extremely similar. In comparison, the proposed method's generated face is the best in terms of all mentioned aspects.

\begin{figure}[t]
    \centering
        \subfigure[WGAN-GP]{
            \includegraphics[width=4.7cm]{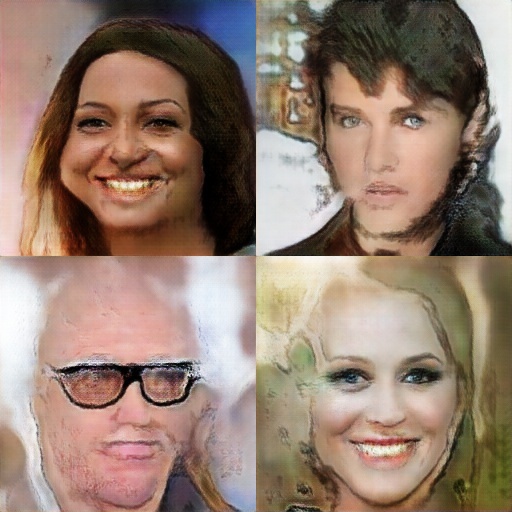}
        }
        \subfigure[S-IntroVAE]{
            \includegraphics[width=4.7cm]{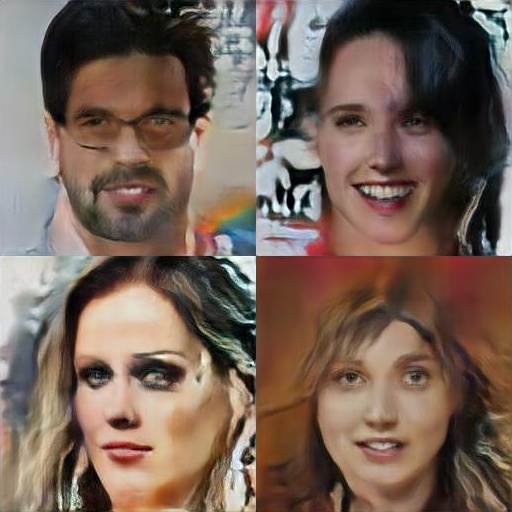}
        }
        \subfigure[Ours]{
            \includegraphics[width=4.7cm]{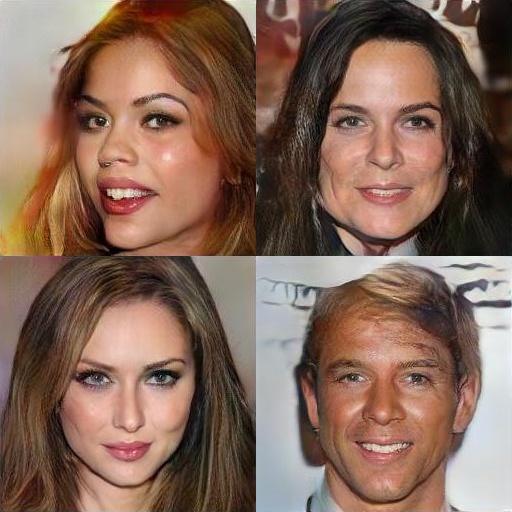}
        } 
    \caption{Image Generation Visual Comparison at CelebA-256 dataset.}
    \label{fig:gen_celeba_256}
    \end{figure}

    Fig. \ref{fig:gen_celeba_256} shows the qualitative comparison of image generation at CelebA-256 dataset. Compared with the CelebA-128 results in Fig. \ref{fig:gen_celeba_128}, we find that both WGAN-GP and S-IntroVAE experience less posterior collapse, unnatural texture, and facial asymmetry. However, both WGAN-GP and S-IntroVAE still have significant facial feature distortion. The ghost artifacts, the edging blur, and the over-smoothed hairs from these methods further degrade the perceptual quality. Compared with WGAN-GP and S-IntroVAE, the proposed method is the best in all mentioned aspects. Due to the scope of this paper, the qualitative result of image generation on other datasets will be in the supplementary material.


    \begin{table}[t]
    \centering
    \begin{minipage}[t]{0.48\textwidth}

        \resizebox{\textwidth}{12mm}{

\begin{tabular}{clllll}
\hline
                    &     & VAE   & IntroVAE & S-IntroVAE & Ours         \\ \hline
\multirow{2}{*}{C1} & KL  & 220.2 & 192.4    & 50.2       & \textbf{3.4} \\
                    & JSD & 110.1 & 56.0     & 16.9       & \textbf{5.6} \\
\multirow{2}{*}{C2} & KL  & 220.3 & 191.1    & 136.5      & \textbf{1.3} \\
                    & JSD & 110.0 & 68.0     & 36.6       & \textbf{4.4} \\
\multirow{2}{*}{C3} & KL  & 220.2 & 64.0     & 46.2       & \textbf{2.0} \\
                    & JSD & 109.8 & 53.0     & 9.6        & \textbf{7.1} \\ \hline
\end{tabular}}

        \label{tab:toy_dataset}
        \caption{2D Toy Dataset 8 Gaussians Score KL$\downarrow$/JSD$\downarrow$ Table}
    \end{minipage}
	\begin{minipage}[t]{0.48\textwidth}
        \centering
    \resizebox{\textwidth}{12mm}{
    \begin{tabular}{llll}
    \hline
               & WGAN-GP & S-IntroVAE & Ours            \\ \hline
    MNIST      & 139.02  & 98.84      & \textbf{96.16}  \\
    CIFAR-10   & 434.11  & 275.20     & \textbf{271.69} \\
    CelebA-128 & 160.53  & 140.35     & \textbf{130.74} \\
    CelebA-256 & 170.79  & 143.33     & \textbf{129.61} \\ \hline
    \end{tabular}}

        \label{tab:fid}
        \caption{Image Generation FID Score$\downarrow$ Table.}
    \end{minipage}
    \end{table}
    

    \subsection{Image Reconstruction}
    In this subsection, we evaluate the proposed method's performance on image reconstruction tasks using benchmark datasets, including MNIST, CIFAR-10, Oxford Building Datasets, CelebA-128, and CelebA-256. The model for comparison is S-IntroVAE, and the evaluation metrics are Peak signal-to-noise ratio (PSNR), Structural Similarity Index (SSIM), and Mean Squared Error (MSE). A higher PSNR, a higher SSIM, and a lower MSE mean better results. 
    
\begin{figure}[t]
    \centering
        \subfigure[Ground truth A]{
            \includegraphics[width=7cm]{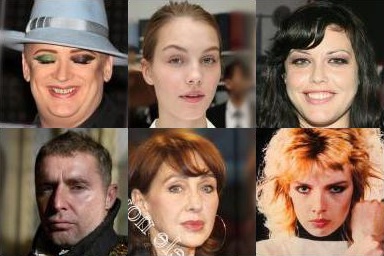}
        }
        \subfigure[S-IntroVAE's reconstruction at A]{
            \includegraphics[width=7cm]{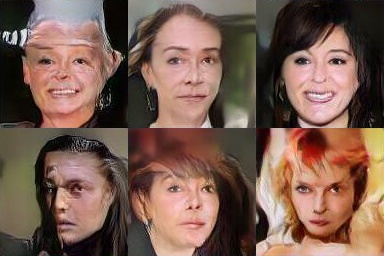}
        }
        \subfigure[Ground truth B]{
            \includegraphics[width=7cm]{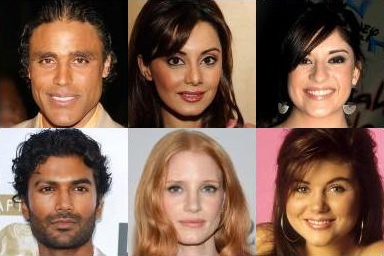}
        } 
        \subfigure[Our's reconstruction at B]{
            \includegraphics[width=7cm]{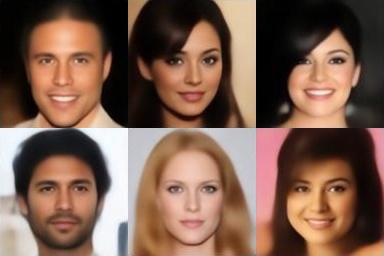}
        } 
    \caption{Image Reconstruction Visual Comparison at CelebA-128 dataset.}
    \label{fig:rec_celeba_128}
    \end{figure}

    Table \ref{tab:PSNR_SSIM_MSE} shows the quantitative comparison of image reconstruction task. For all except CelebA-256, our method has the best PSNR, SSIM, and MSE. For the CelebA-256 dataset, our method has the second-best SSIM but the best for both PSNR and MSE. Fig. \ref{fig:rec_celeba_128} shows the qualitative comparison of image reconstruction at CelebA-128 dataset. S-IntroVAE fails to faithfully reconstruct the facial features, facial expressions, and skin textures. The reconstructed image also contains significant edging blur, defects, and artifacts, which significantly distort the perceptual quality.

    The proposed method is much closer to the ground truth regarding contrast, exposure, color, edge information, and facial details. In short, our approach surpasses S-IntroVAE by a large margin in image reconstruction with the CelebA-128 dataset. 
    

    
    
\begin{table}[t]
    \centering
    \begin{tabular}{ccccccc}
    \hline
    \multicolumn{1}{l}{} & \multicolumn{2}{c}{PSNR}     & \multicolumn{2}{c}{SSIM}        & \multicolumn{2}{c}{MSE}         \\
    \multicolumn{1}{l}{} & S-IntroVAE & Ours            & S-IntroVAE     & Ours           & S-IntroVAE     & Ours           \\ \hline
    MNIST                & 20.282     & \textbf{21.014} & 0.885          & \textbf{0.898} & 0.011          & \textbf{0.009} \\
    CIFAR-10             & 19.300     & \textbf{19.445} & 0.599          & \textbf{0.620} & \textbf{0.019} & \textbf{0.019} \\
    Oxford               & 15.372     & \textbf{20.168} & 0.348          & \textbf{0.604} & 0.049          & \textbf{0.013} \\
    CelebA-128           & 17.818     & \textbf{22.924} & 0.561          & \textbf{0.801} & 0.018          & \textbf{0.006} \\
    CelebA-256           & 22.422     & \textbf{23.156} & \textbf{0.790} & 0.758          & 0.007          & \textbf{0.006} \\ \hline
    \end{tabular}
    \label{tab:PSNR_SSIM_MSE}
    \caption{Image Reconstruction PSNR$\uparrow$/SSIM$\uparrow$/MSE$\downarrow$ Score Table}
\end{table}

        
    
    \subsection{Training Stability}
    \begin{figure}[h]
        \centering
        \subfigure[IntroVAE]{
        \includegraphics[width=4cm]{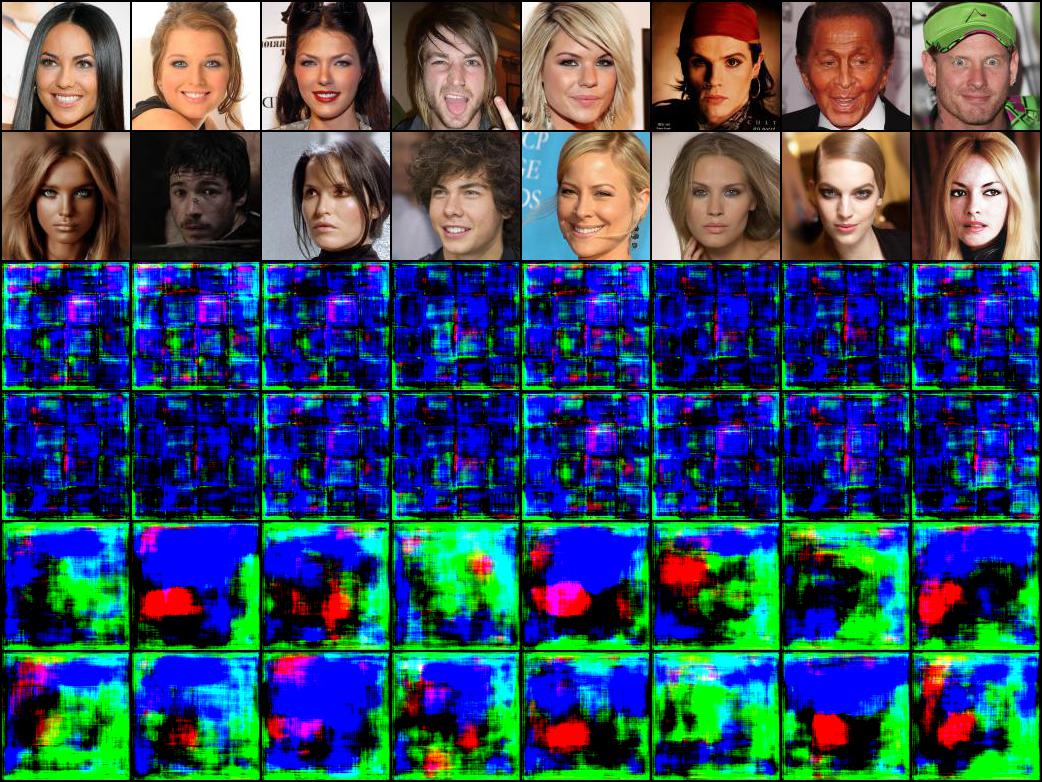}
        \includegraphics[width=4cm]{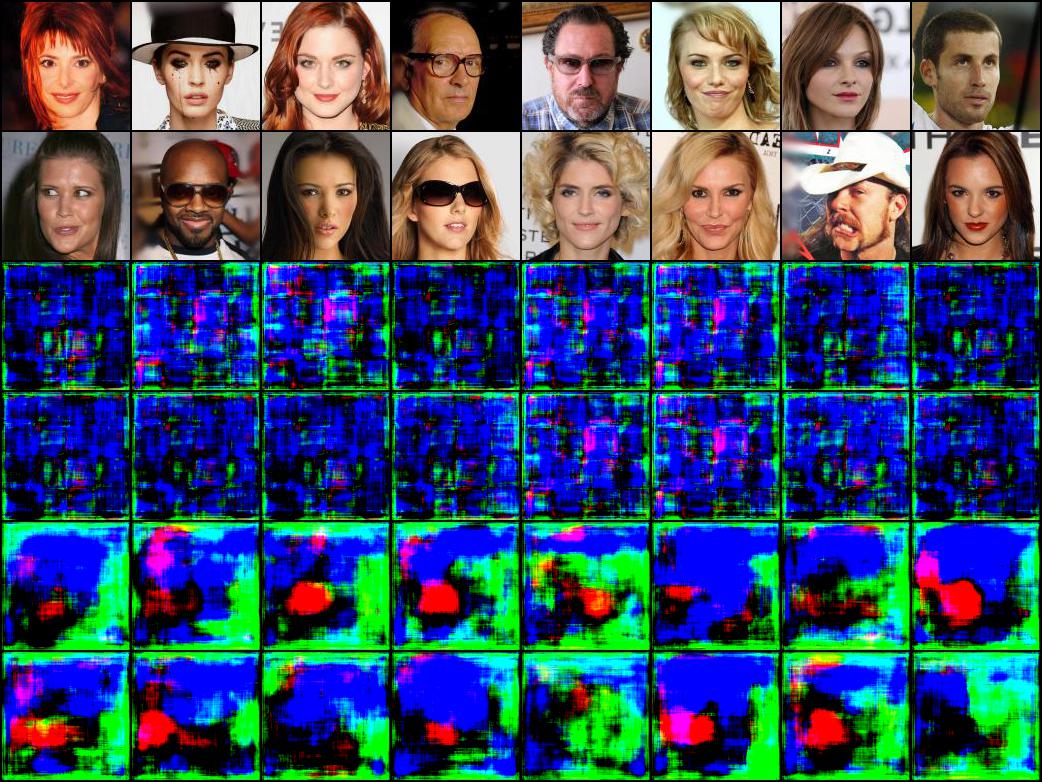}
        \includegraphics[width=4cm]{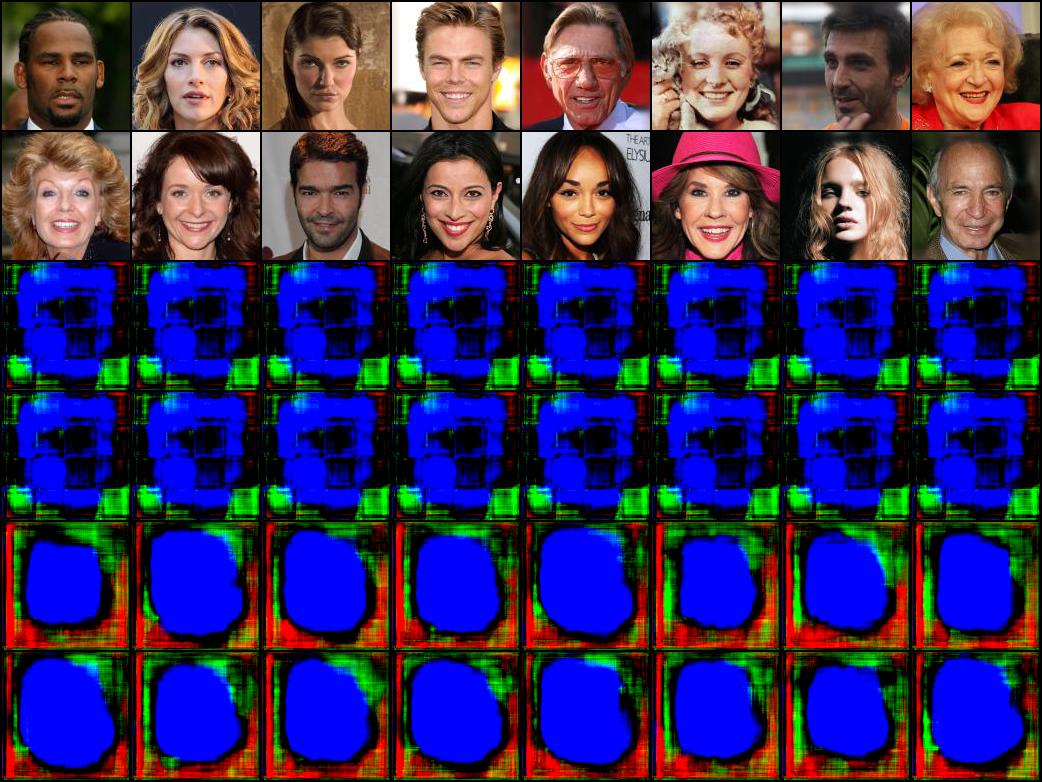}
        }
        
        \subfigure[S-IntroVAE]{
        \includegraphics[width=4cm]{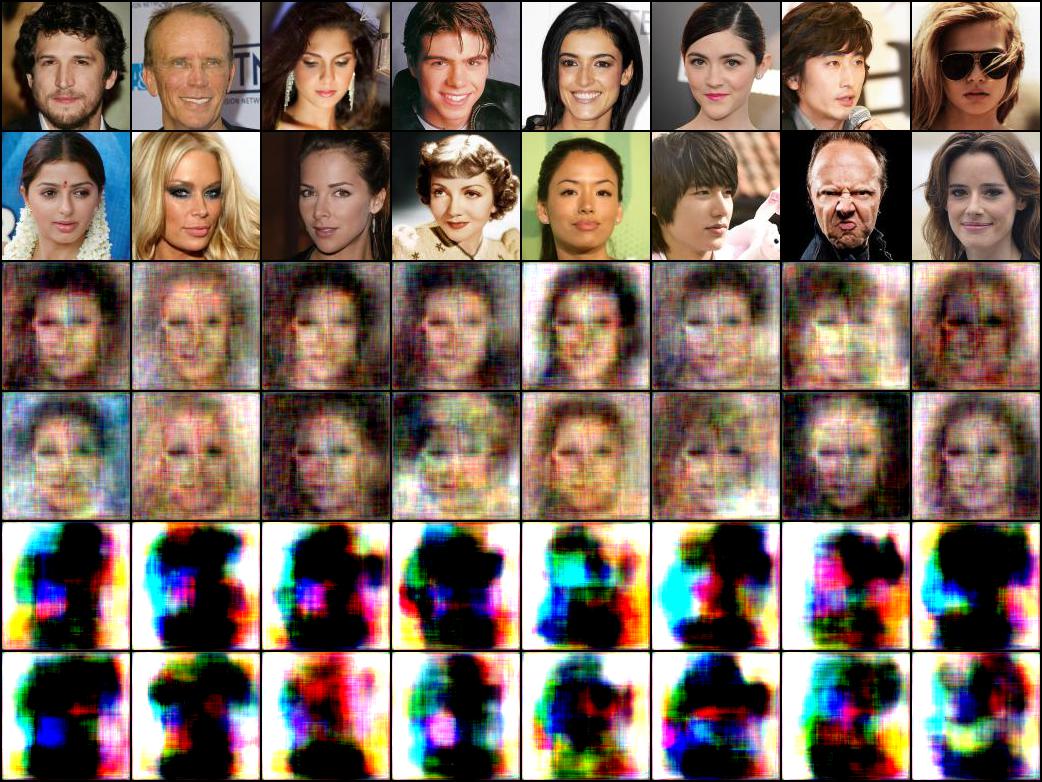}
        \includegraphics[width=4cm]{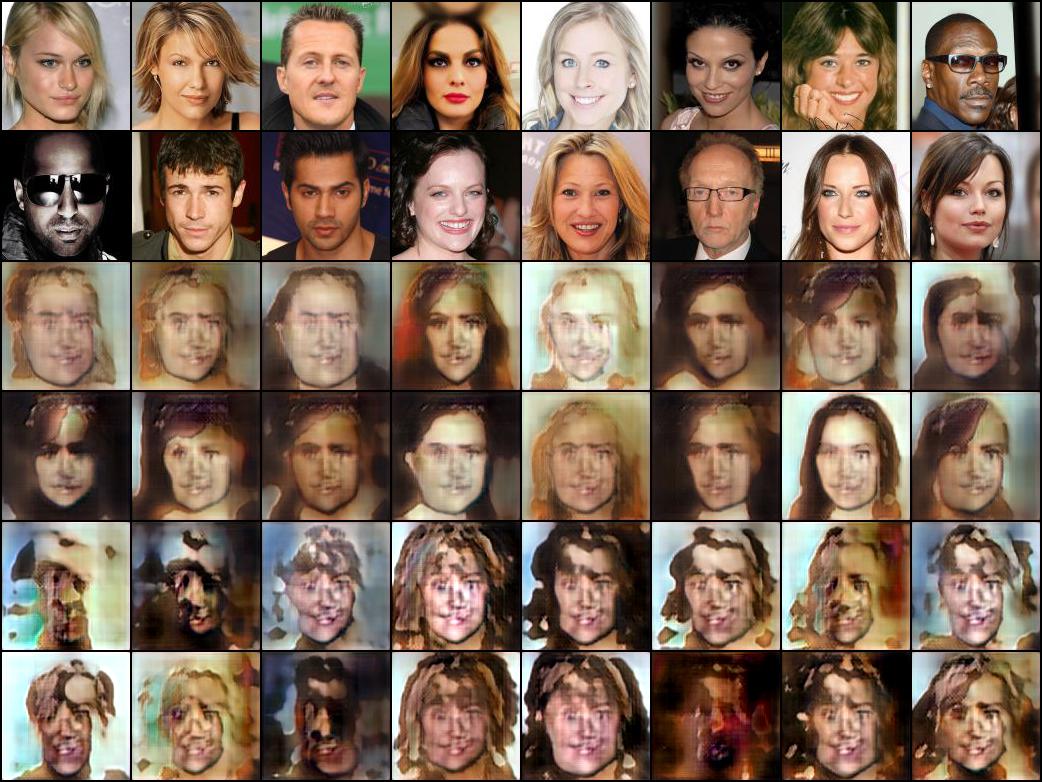}
        \includegraphics[width=4cm]{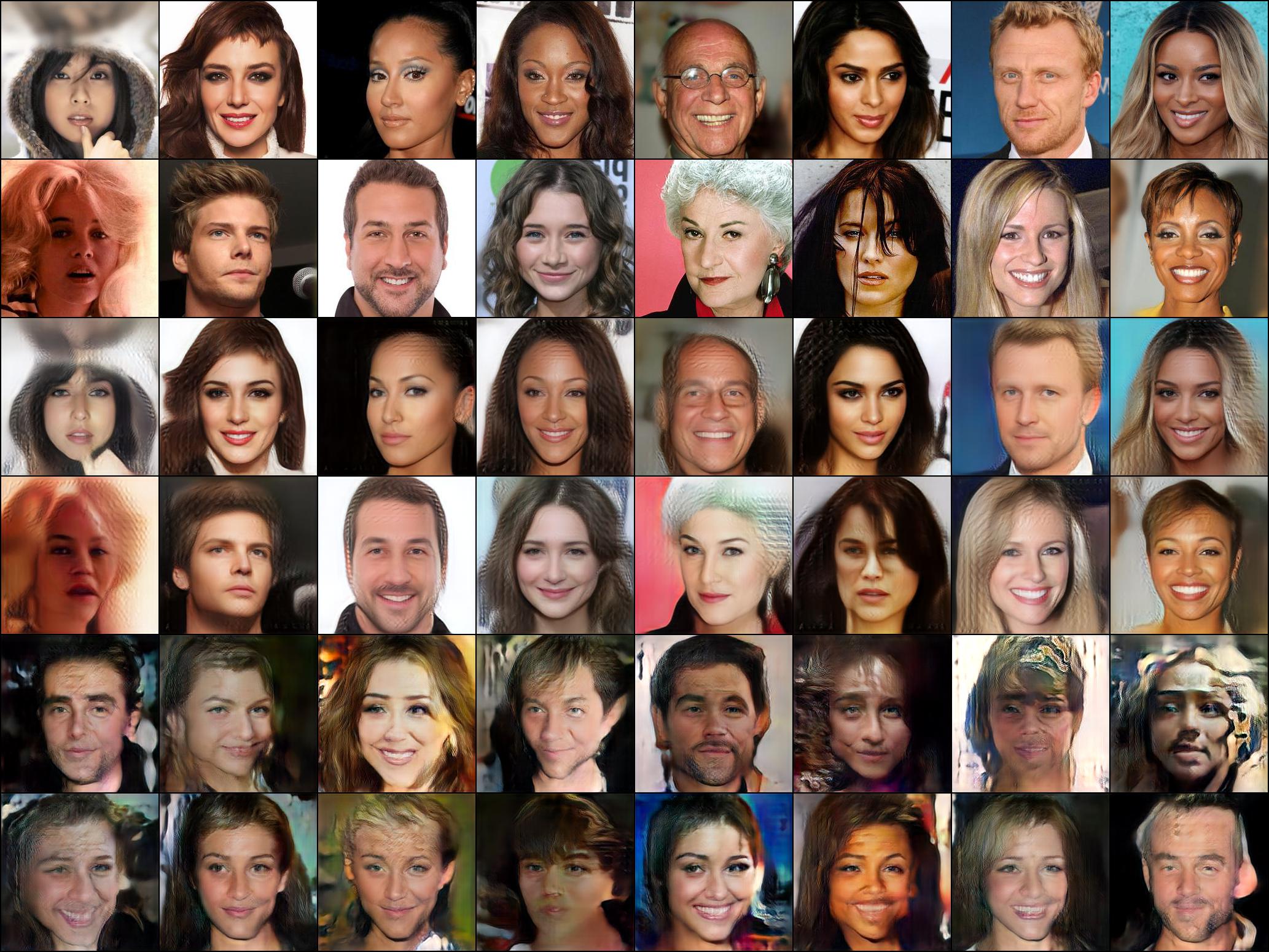}
        }
        
        \subfigure[AS-IntroVAE (Ours)]{
        \includegraphics[width=4cm]{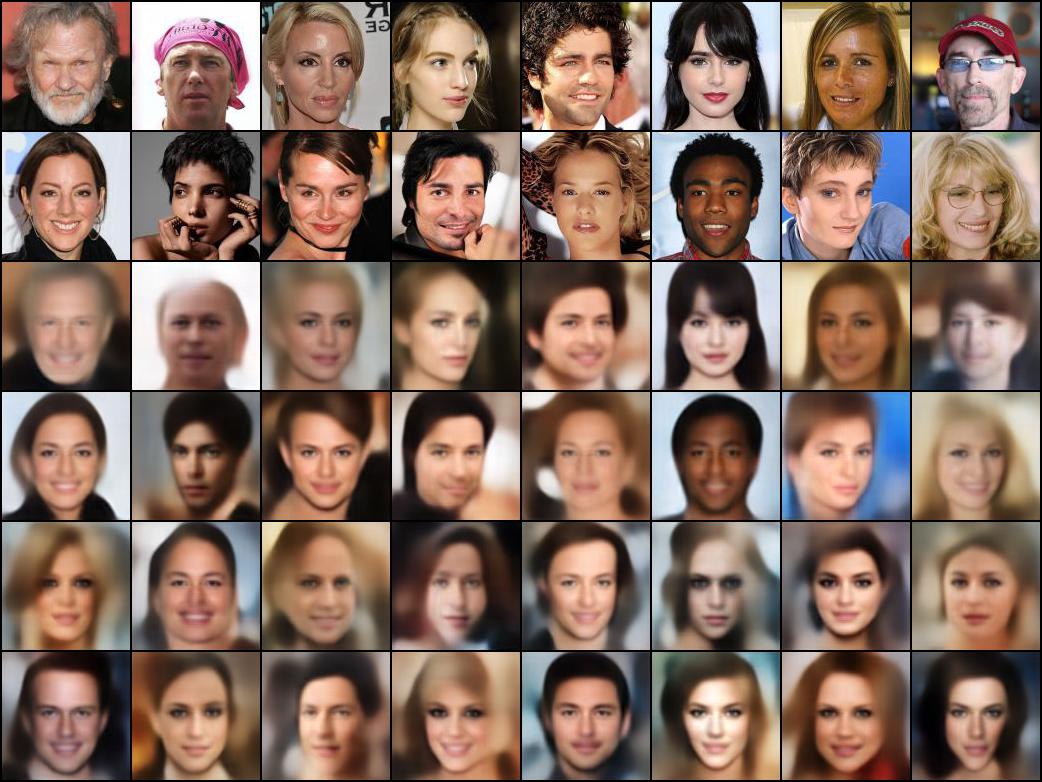}
        \includegraphics[width=4cm]{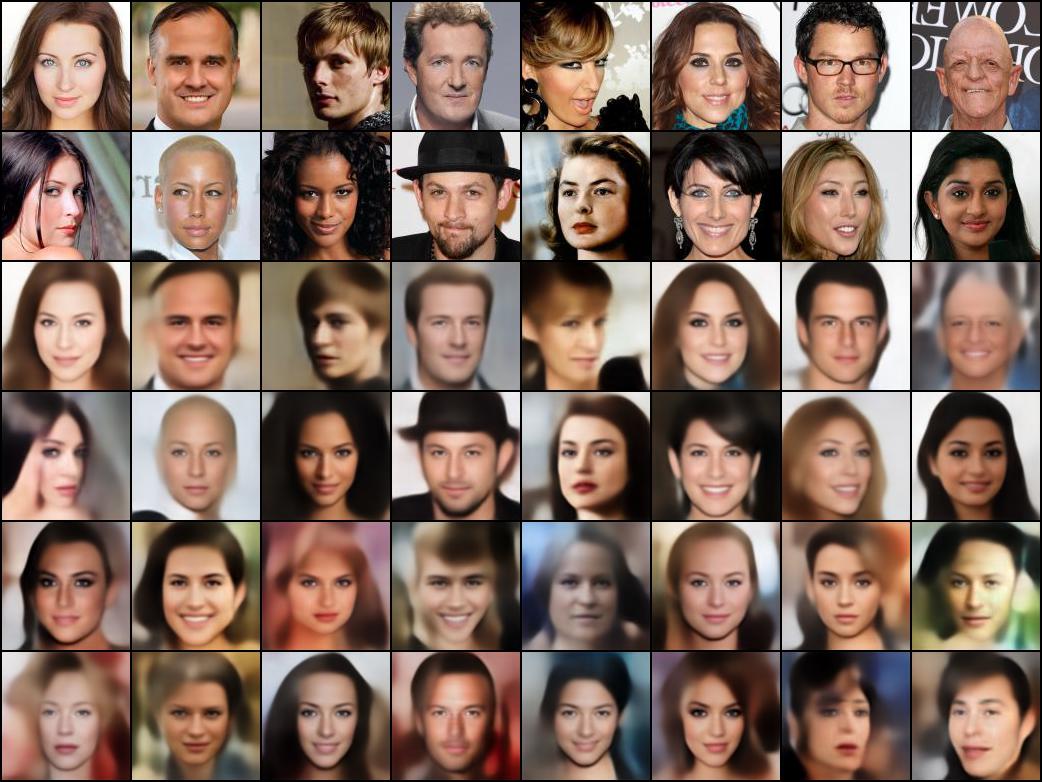}
        \includegraphics[width=4cm]{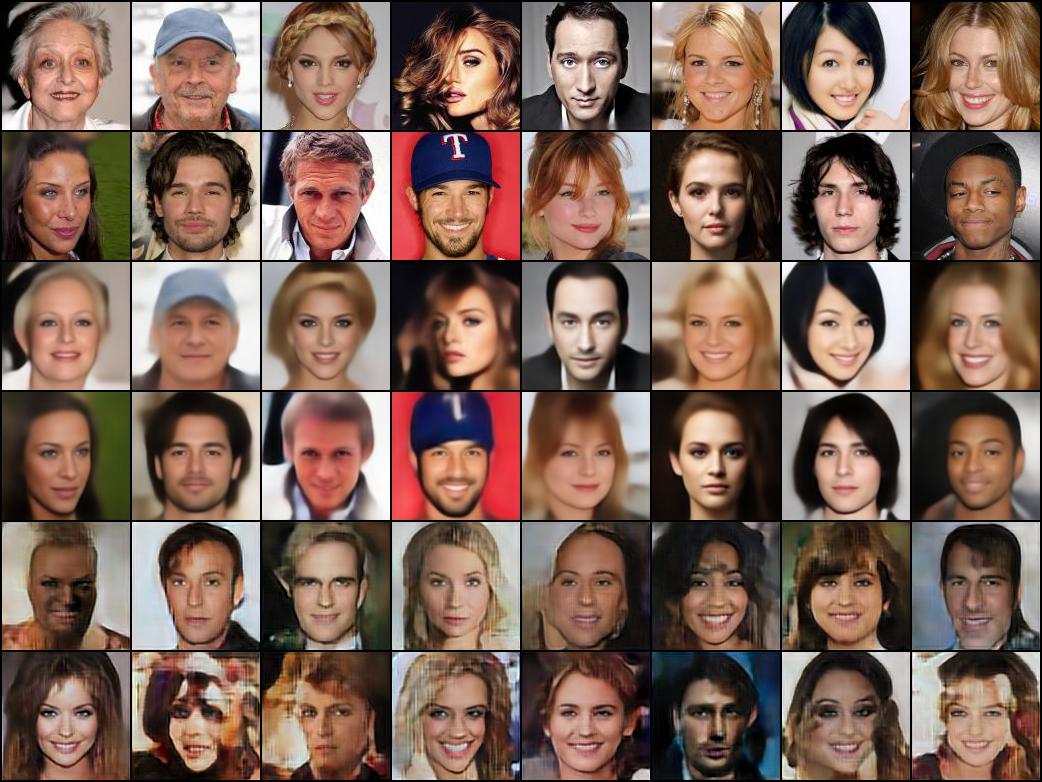}
        }
        
        \caption{The training stability visual comparison at CelebA-128 dataset. From left to right panel: 10 epoch, 20 epoch, 50 epoch. For each image grid, the first and the second row are real images, the third and fourth rows are reconstructed images, and the fifth and sixth rows are generated images. Zoom in for a better view.}
        \label{fig:train_stability}
    \end{figure}
    
    Fig. \ref{fig:train_stability} shows the training stability visual comparison at CelebA-128 dataset. We find that IntroVAE fails in image reconstruction and image generation tasks, even if we train the model using its recommended hyperparameters. IntroVAE's reconstructed images at a specific epoch are almost homogeneous: a mixture of blue and green clouds with little semantic information. 
        
    
    We also find that S-IntroVAE has split performances in the early stage (10 \& 20 epoch) and the later stage (50 epoch). In the early stage, the reconstructed images contain a loss of blur, defects, and artifacts, whereas the generated images have distorted facial features and a significant amount of unnatural artifacts. In the later stage, the quality of both tasks has improved. However, the generated and reconstructed images still contain many defects and edging blur.

    In comparison, the proposed method has quick learning convergence in the early stage (10 \& 20 epoch) and maintains excellent training stability in the later stage (50 epoch). The reconstructed images are faithfully aligned with the original images, whereas the generated images have superb perceptual quality.


\section{Conclusion}

This paper introduces Adversarial Similarity Distance Introspective Variational Autoencoder (AS-IntroVAE), a new introspective approach that can faithfully address the posterior collapse and the vanishing gradient problem. Our theoretical analysis rigorously illustrated the advantages of the proposed Adversarial Similarity Distance (AS-Distance). Our empirical results exhibited compelling quality, diversity, and stability in image generation and construction tasks. In the future, we hope to apply the proposed AS-IntroVAE to high resolution (e.g., 1024 $\times$ 1024) image synthesis. We also hope to extend AS-IntroVAE to reinforcement learning, self-supervised learning tasks with detection-driven (\cite{zheng2021deblur}) and segmentation-driven (\cite{zheng2022sapnet}) techniques, and medical image analysis (\cite{lu2022unsupervised}).



\bibliography{acml22}

\appendix

\section{supplementary material}
\subsection{Introduction}
In this supplementary material, we first show experiment results with different weights for Adversarial Similarity Distance (AS-Distance) and KL Divergence, and then proceeds to more visual comparisons on image generation tasks on various benchmark dataset.

\subsection{AS-Distance and KL Divergence}
In this section, we use a visual comparison of image generation and image reconstruction tasks to show that the following hyperparameter combinations are worse than the weight annealing method introduced in the paper. The hyperparameter combinations are (1) AS-IntroVAE with a weight of 1.0 for AS-Distance and 0 for KL Divergence and (2) AS-IntroVAE with a weight of 0.5 for both AS-Distance and KL Divergence. 

\begin{figure}[h]
    \centering
    \includegraphics[width=16cm]{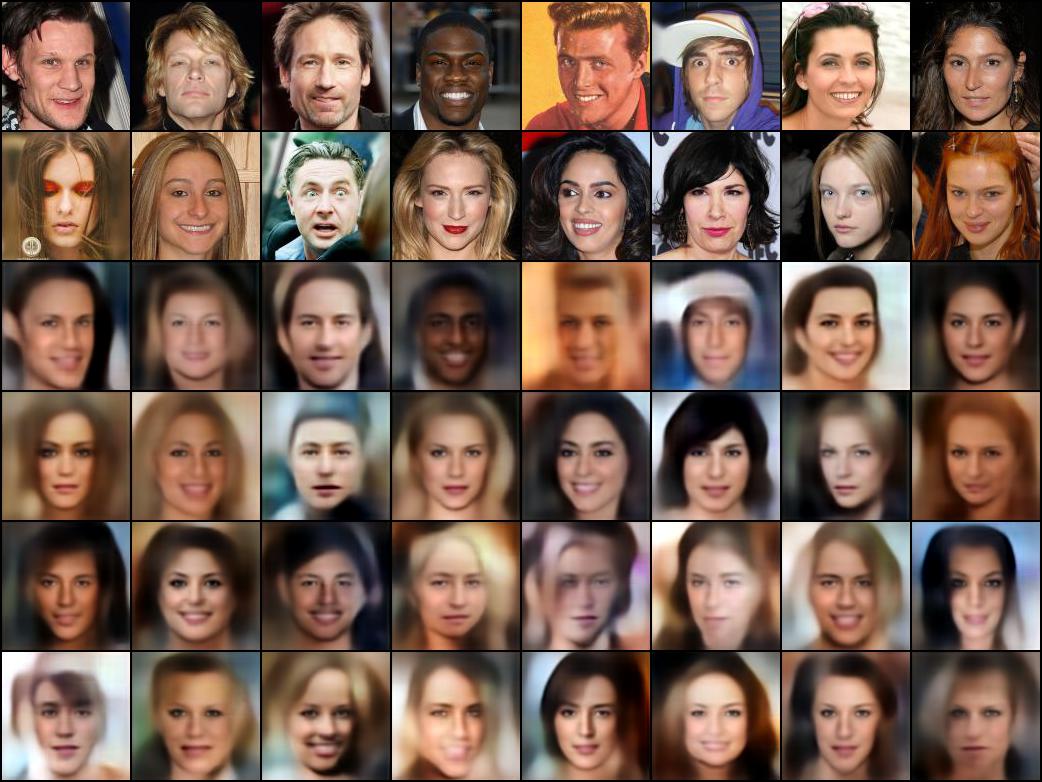}
    \caption{AS-IntroVAE performance at CelebA-128, using only AS-Distance and no KL divergence. The upper/middle/bottom two row refer to real/reconstructed/generated images. We can see that the images are over-smoothed and looks blurry without the help of KL divergence.}
\end{figure}



\begin{figure}[h]
    \centering
    \includegraphics[width=16cm]{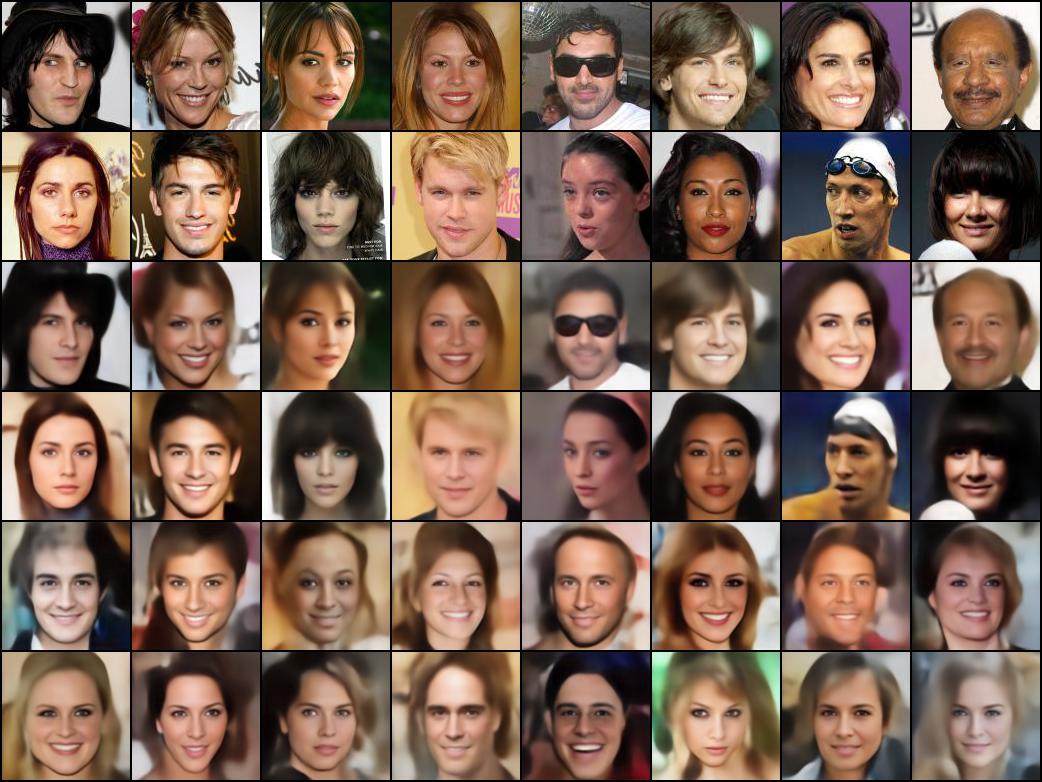}
    \caption{S-IntroVAE performance at CelebA-128, when the weight for KL divergence and AS-Distance are both 0.5. The upper/middle/bottom two rows refer to real/reconstructed/generated images. We can see that the images are with significant blur. }
\end{figure}

\begin{figure}[h]
    \centering
    \includegraphics[width=16cm]{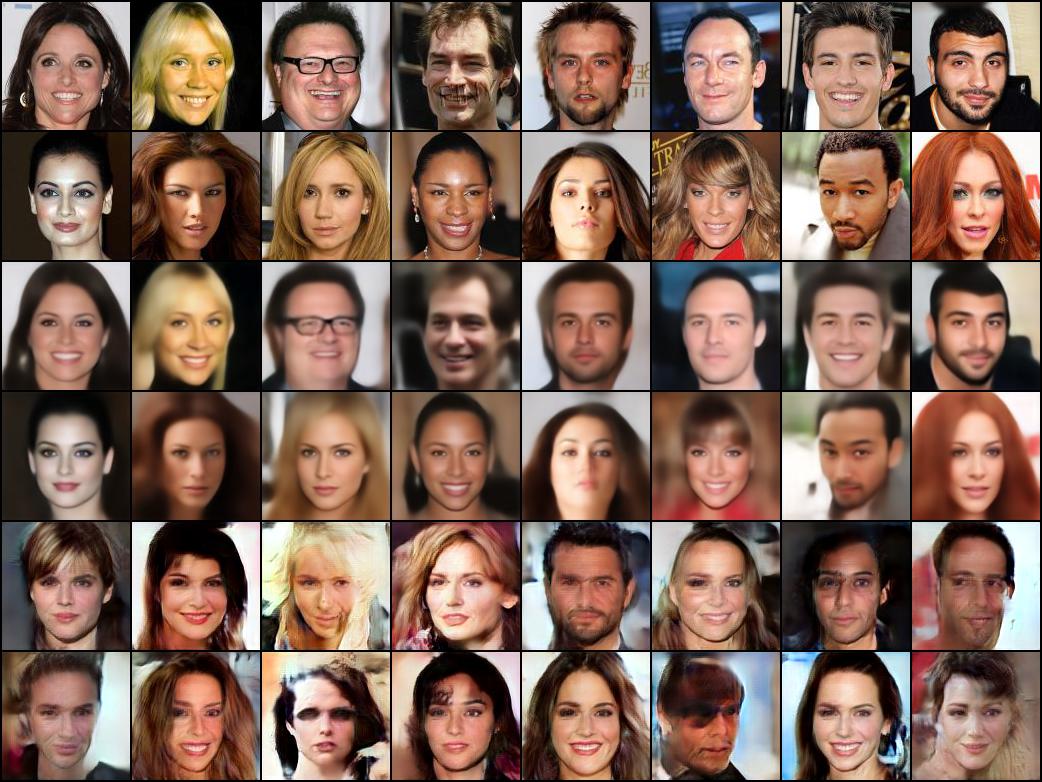}
    \caption{AS-IntroVAE performance at CelebA-128, when the weight for KL divergence and AS-Distance are both 0.5. The upper/middle/bottom two rows refer to real/reconstructed/generated images. From this figure and the figure above, we note that different images display different levels of sharpness and blur. Therefore, we conclude that this hyperparameter combination causes the model to have unstable training and fluctuating performances.}
\end{figure}

\clearpage
\section{Visual Comparison for Image Generation}
This section shows the additional visual comparison for image generation tasks. Specifically, we display the results on four datasets, including CelebA-128, CelebA-256, MNIST, and CIFAR10. For each dataset, we randomly select 16 images from each model's output dataset. In each figure, the upper left images are from AS-IntroVAE, the upper right images are from S-IntroVAE, and the bottom images are from WGAN-GP. Note that 
\begin{figure}[h]
    \centering
    \includegraphics[width=7.5cm]{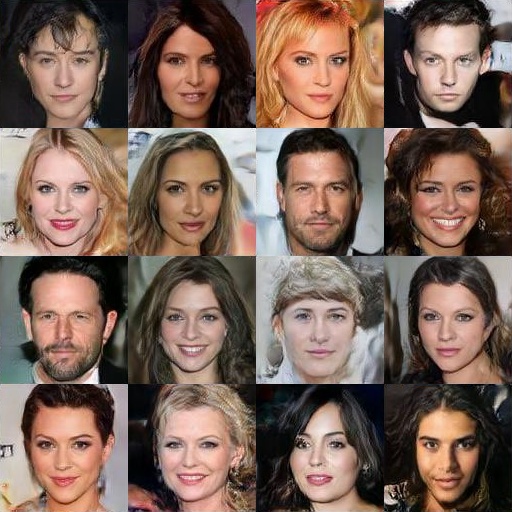}
    \includegraphics[width=7.5cm]{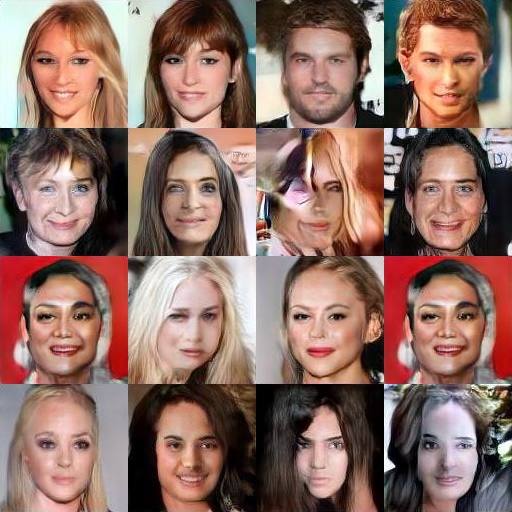}
    \includegraphics[width=7.5cm]{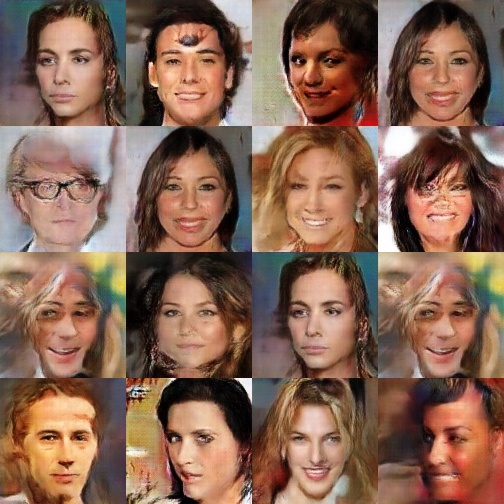}
    \caption{Image generation visual comparisons at CelebA-128 dataset (resolution: 128 $\times$ 128).}
\end{figure}

\begin{figure}[h]
    \centering
    \includegraphics[width=7.5cm]{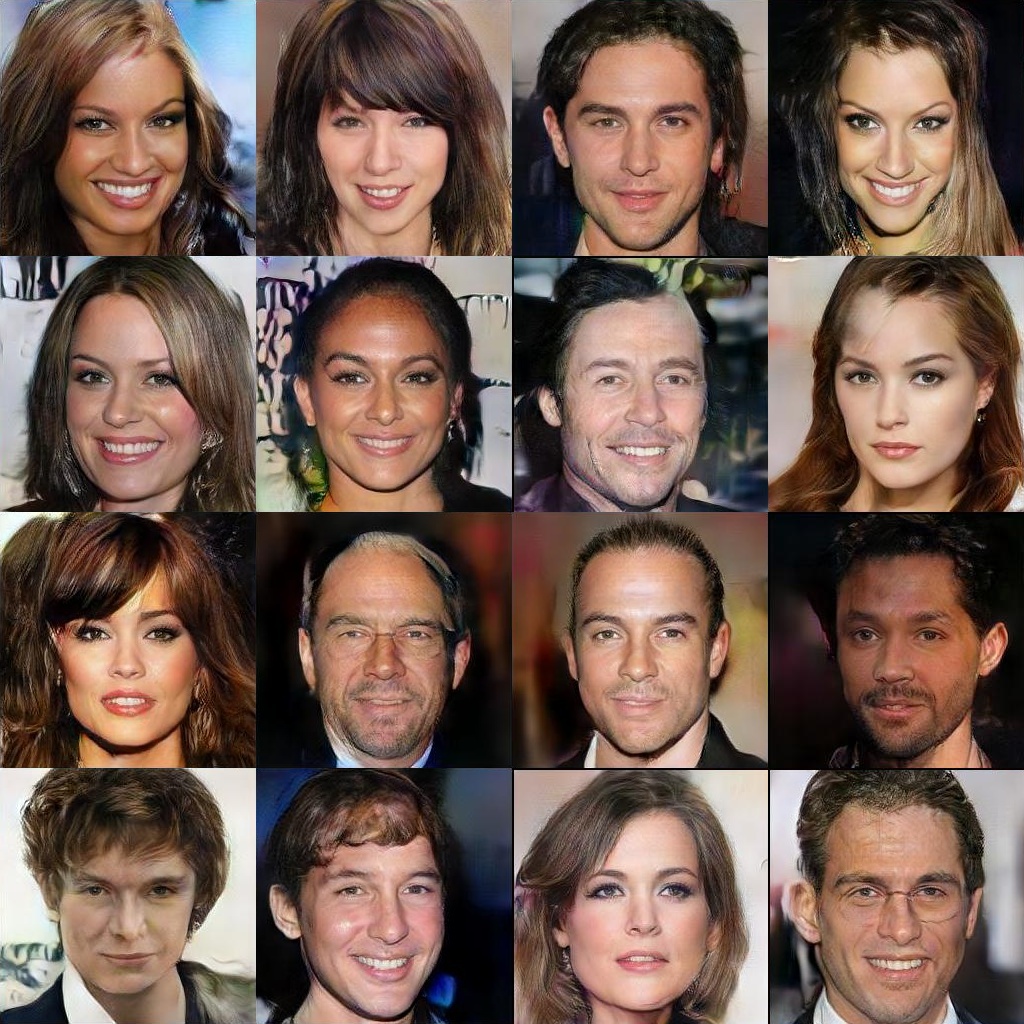}
    \includegraphics[width=7.5cm]{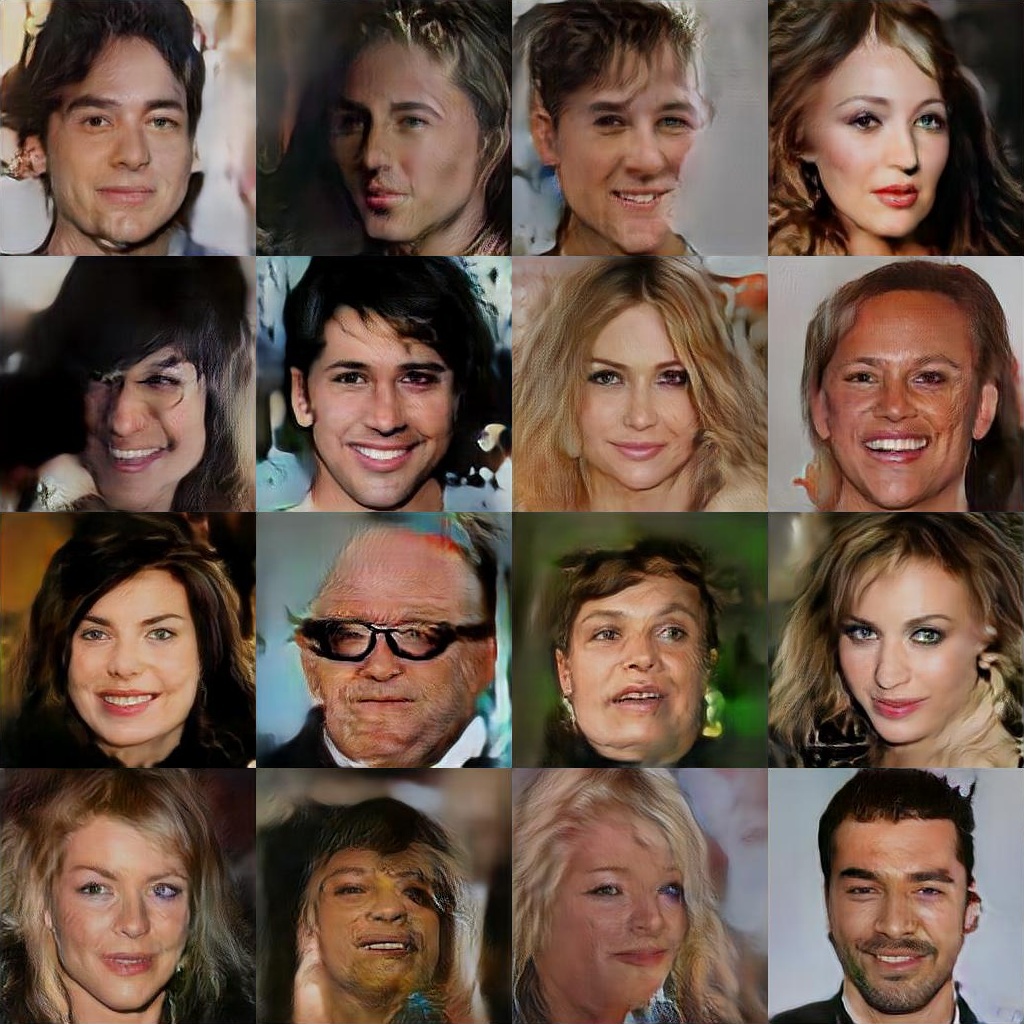}
    \includegraphics[width=7.5cm]{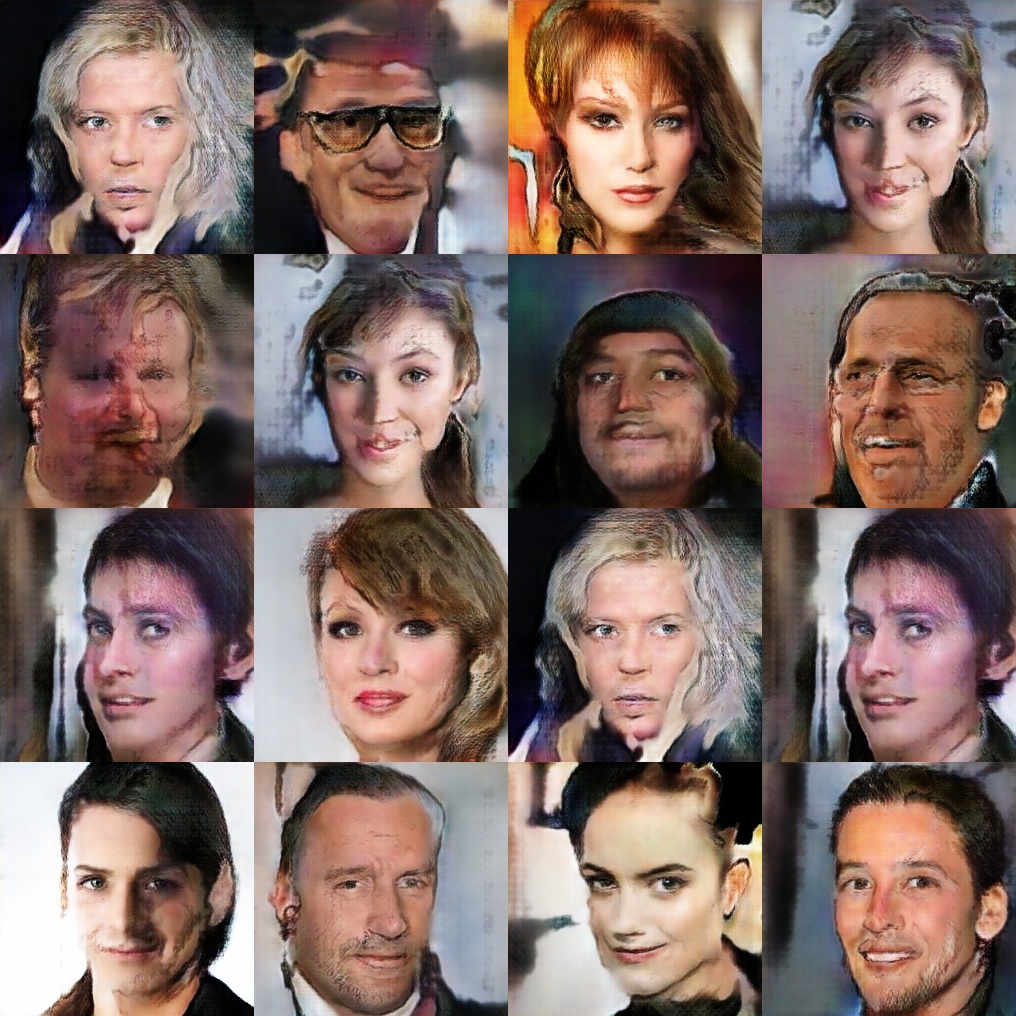}
    \caption{Image generation visual comparisons at CelebA-256 dataset (resolution: 256 $\times$ 256).}
\end{figure}

\begin{figure}[h]
    \centering
    \includegraphics[width=7.5cm]{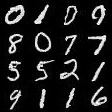}
    \includegraphics[width=7.5cm]{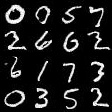}
    \includegraphics[width=7.5cm]{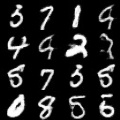}
    \caption{Image generation visual comparisons at MNIST dataset (resolution: 28 $\times$ 28).}
\end{figure}

\begin{figure}[h]
    \centering
    \includegraphics[width=7.5cm]{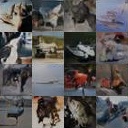}
    \includegraphics[width=7.5cm]{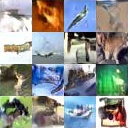}
    \includegraphics[width=7.5cm]{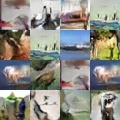}
    \caption{Image generation visual comparisons at CIFAR10 dataset (resolution: 32 $\times$ 32)}
\end{figure}

\clearpage
    \begin{figure}[h]
        \centering
        \includegraphics[width=3cm]{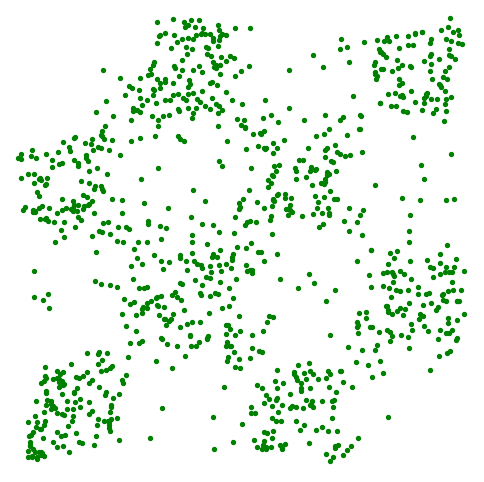}
        \includegraphics[width=3cm]{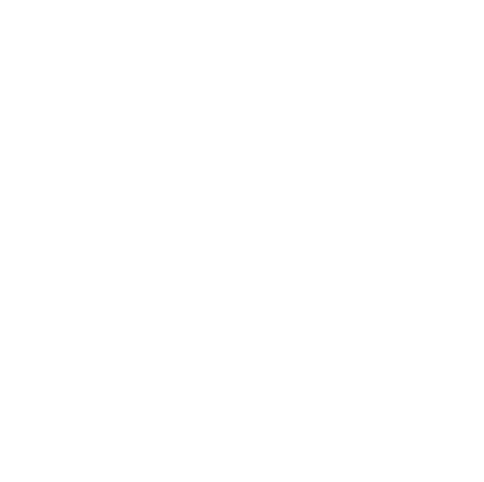}
         \includegraphics[width=3cm]{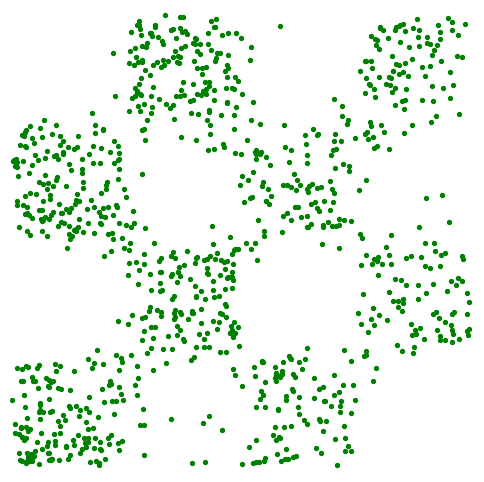}
        \includegraphics[width=3cm]{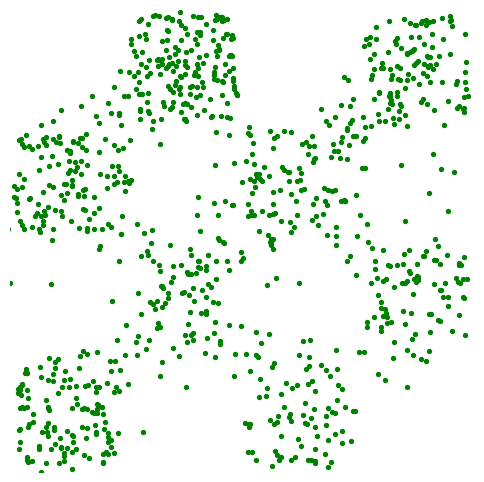}

        \includegraphics[width=3cm]{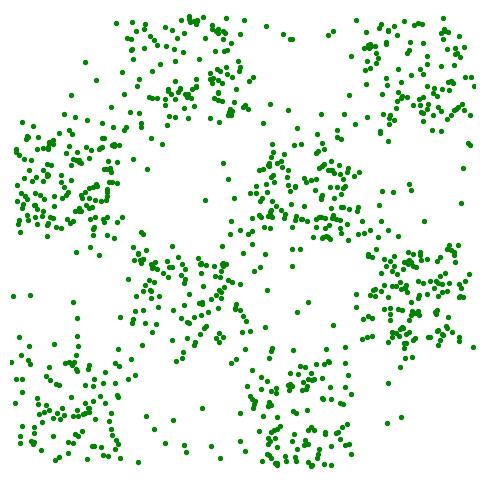}
         \includegraphics[width=3cm]{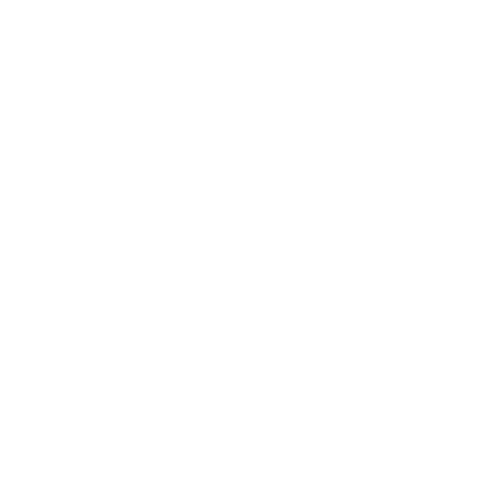}
         \includegraphics[width=3cm]{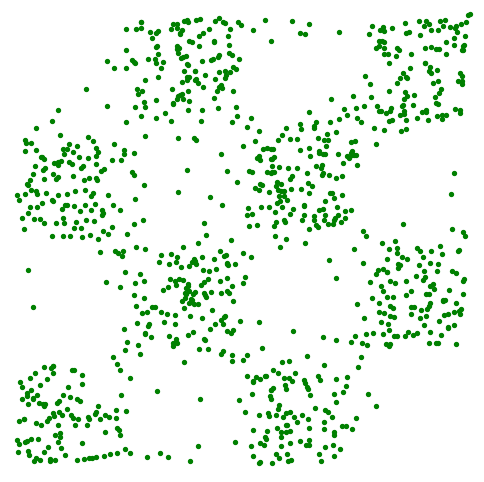}
         \includegraphics[width=3cm]{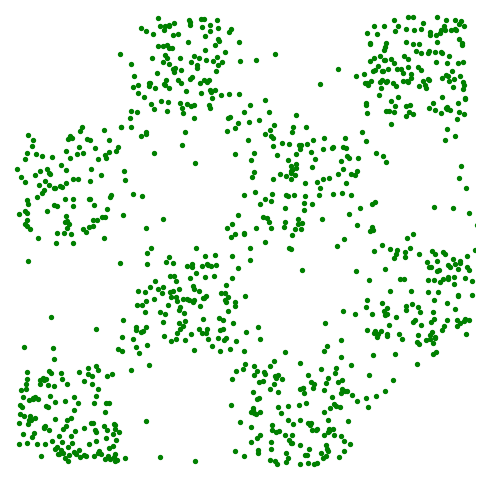}

        \includegraphics[width=3cm]{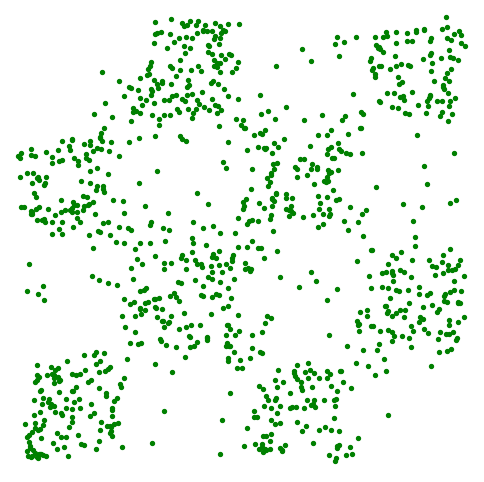}
        \includegraphics[width=3cm]{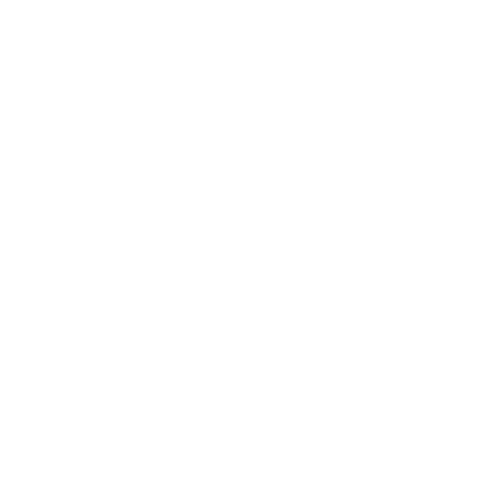}
        \includegraphics[width=3cm]{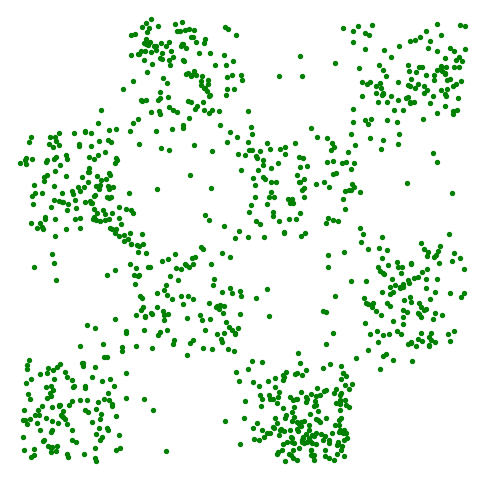}
        \includegraphics[width=3cm]{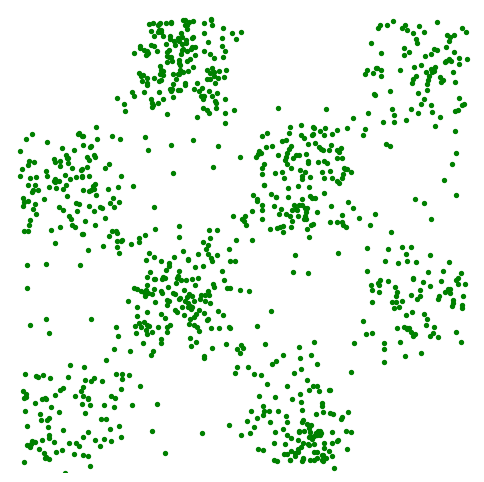}

        \caption{Visual Comparison on 2D Toy Dataset Checkerboard. From top to bottom row: results with different hyperparameters. From left to right column: VAE, IntroVAE, S-IntroVAE, Ours. The results show that AS-IntroVAE has a slight advantage over S-IntroVAE in terms of point clustering and centroid convergence.}
        \label{fig:Checkerboard}
    \end{figure}

    \begin{table}[h]
        \centering
        \begin{tabular}{clllll}
    \hline
                        &     & VAE   & IntroVAE & S-IntroVAE & Ours         \\ \hline
    \multirow{2}{*}{C1} & KL  & 22.1 & NaN    & 20.7       & \textbf{20.4} \\
                        & JSD & 10.8 & --    & \textbf{9.6}      & \textbf{9.6} \\
    \multirow{2}{*}{C2} & KL  & 21.2 & NaN   & 21.0      & \textbf{20.6} \\
                        & JSD & 9.9 & --      & 10.0      & \textbf{9.6} \\
    \multirow{2}{*}{C3} & KL  & 21.7 & NaN    & 21.2       & \textbf{20.9} \\
                        & JSD & 10.7 & --      & 10.3        & \textbf{9.9} \\ \hline
    \end{tabular}
            \caption{2D Toy Dataset Checkerboard  KL$\downarrow$/JSD$\downarrow$ Score Table. The Table shows that the proposed AS-IntroVAE has the best score for KL and JSD under all hyperparameter combinations. }
            \label{tab:my_label}
        \end{table}

\end{document}